\documentclass{article}
\usepackage{graphicx,emulateapj}

\def\myputfigure#1#2#3#4#5%
{\vskip#5pt\makebox[0pt]{\hskip#2in
\includegraphics[width=#3\textwidth]{#1}}\vskip#4pt\hfill}

\def\gap{\;\rlap{\lower 2.5pt
 \hbox{$\sim$}}\raise 1.5pt\hbox{$>$}\;}
\def\lap{\;\rlap{\lower 2.5pt
   \hbox{$\sim$}}\raise 1.5pt\hbox{$<$}\;}
\def\gsim{\;\rlap{\lower 2.5pt
 \hbox{$\sim$}}\raise 1.5pt\hbox{$>$}\;}
\def\lsim{\;\rlap{\lower 2.5pt
   \hbox{$\sim$}}\raise 1.5pt\hbox{$<$}\;}

\def\spose#1{\hbox to 0pt{#1\hss}}
\def\lta{\mathrel{\spose{\lower 3pt\hbox{$\mathchar''218$}}
     \raise 2.0pt\hbox{$\mathchar''13C$}}}
\def\gta{\mathrel{\spose{\lower 3pt\hbox{$\mathchar''218$}}
     \raise 2.0pt\hbox{$\mathchar''13E$}}}
\newcommand{\beq}{\begin{equation}}
\newcommand{\eeq}{\end{equation}}
\newcommand{\be}{\begin{equation}}
\newcommand{\ee}{\end{equation}}

\lefthead{VERDE, HAIMAN \& SPERGEL}
\righthead{GALAXY CLUSTER SCALING RELATIONS}

\begin{document}
\title{Are Clusters Standard Candles? Galaxy Cluster Scaling Relations With the Sunyaev-Zeldovich Effect}
\author{Licia Verde}
\affil{Princeton University Observatory, Princeton, NJ 08544, USA and\\ Dept. of
Physics and Astronomy, Rutgers University, 136 Frelinghuysen Road, Piscataway NJ 08854-8019 \\}
\author{Zolt\'an Haiman\altaffilmark{1} and David N. Spergel}
\affil{Princeton University Observatory, Princeton, NJ 08544, USA\\ 
lverde,zoltan,dns@astro.princeton.edu}
\altaffiltext{1}{Hubble Fellow}

\begin{abstract}
An extensive sample of galaxy clusters will be available in the 
coming years, detected through their Sunyaev--Zeldovich effect (SZE). 
We use a semi--analytic model to study the scientific yield of 
combining SZE data with X-ray and optical follow-up observations.
If clusters at a given redshift $z_o$ can be identified with virialized, 
spherical halos, they populate a well--defined "fundamental plane" (FP) 
in the parameter space of the three observables: virial temperature ($T$), 
total Sunyaev--Zeldovich flux decrement ($\Delta S_\nu$), and angular 
size ($\theta$).  The location and orientation of the FP, 
as well as its redshift--evolution, are sensitive to both the internal 
evolution of clusters, and to the underlying cosmological parameters.  
We show that if clusters are not standard candles (e.g. due to feedback, 
or energy injection), then this can be inferred from the FP. Likewise, 
we study the dependence of the FP on the cosmological parameters $h,\sigma_8$, 
and $\Omega_0$,  and quantify constraints on these parameters.  
We also show that in the absence of any non--gravitational effects, 
the scatter in the the ($\Delta S_\nu-T$) plane is significantly smaller 
than in either the ($\theta-T$) or the ($\theta-\Delta S_{\nu}$) planes.
As a result, the ($\Delta S_\nu-T$) relation can be an exceptionally
sensitive probe of both cluster physics and cosmological parameters,
and a comparison of the amount of scatter in these three scaling relations
will serve as a test of the origin (cosmological vs. stochastic) of 
the scatter.
\end{abstract}
\keywords{Cosmology: theory, cosmological parameters, galaxies: clusters: general}

\section{Introduction}
\label{sec:intro}

Galaxy clusters provide a uniquely useful probe of the fundamental
cosmological parameters.  The observed abundance of nearby clusters
constrains the amplitude $\sigma_{\rm 8}$ and slope $n$ of the
primordial power spectrum on cluster scales
(\cite{Evr89,HenArn91,BC92,WEF93,ECF96,Pierpaoli00}), while the
cluster mass function (\cite{Lil92,BC93,Vianaliddle96}), and its
redshift--evolution (\cite{OukBla92,FB98,BB98,Vianaliddle99,Willick00}), places
useful constrains on the density parameters $\Omega_0$ and
$\Omega_\Lambda$, as well as on the equation of state of the dark
energy component (\cite{WS98,Haimanetal00}).

The analyses above require an understanding of the internal physics of
galaxy clusters, and the extension of the physical properties of the
local cluster sample towards higher redshifts.  In other words, for
cosmological studies, one would ideally like to use clusters as
``standard candles''. The central assumptions that make this possible
are that clusters are virialized objects at characteristic background  densities
scaling with redshift as $\rho\propto(1+z)^3$. Under these
assumptions, the temperatures, masses and redshifts of clusters are
related by the virial theorem.  As demonstrated recently by Haiman,
Mohr \& Holder (2001), the use of large future galaxy cluster surveys
for cosmological studies is likely to be limited by the validity of
these assumptions, rather than by statistical uncertainties.

Galaxy clusters exhibit various useful scaling relations, such as
those between mass and temperature ($M-T$; e.g., \cite{MME99}),
luminosity and temperature ($L-T$; e.g., \cite{ENF98}), or size and
temperature ($R-T$; e.g., \cite{Mohrevrard97}).  Analyses of these,
and other, similar scalings laws have yielded insight into the
physical nature of clusters. In addition, the scaling relations and
their scatter also contain useful cosmological information
(\cite{VKMB}).  The temperature of the hot intra--cluster medium (ICM)
is a uniquely direct measure that can be accurately determined from
X--ray spectra.  On the other hand, the (X--ray) luminosity is
sensitive to the density profiles near the cluster core, and the
measurements of the masses (from gravitational lensing, or from the
velocity dispersion of member galaxies) and radii (e.g., from X--ray
isophotes) typically suffer from systematic uncertainties and are thus
less robust.

A direct probe of the hot gas in galaxy clusters is provided by the
so--called Sunyaev--Zeldovich effect (SZE, \cite{SZ80}). Cosmic
microwave background (CMB) photons interact with the hot ionized
intra--cluster gas along their path, distorting the CMB spectrum. In
the Rayleigh--Jeans regime, the distortion results in a ``decrement''
(SZD) of the CMB temperature that depends only on the optical depth to
Compton scattering and the cluster temperature, but not on the
redshift of the cluster.  While the SZE has currently been measured
only in a handful of clusters, future SZE surveys can detect rich
clusters up to arbitrary redshifts; for example, a recently initiated
interferometric survey (\cite{Holderetal00}), covering 12 deg$^2$,
will detect hundreds of clusters below redshift $z=3$.

The SZE can be characterized in two ways: by its "surface brightness" (i.e. the
value of the SZ decrement per unit solid angle) near the central region of the
cluster, or by its total flux decrement (i.e. the decrement integrated over the
whole solid angle of the cluster).  While at present SZE studies have mainly
used the central SZ decrement, here we propose to use the total SZ flux
decrement. This quantity should be a more robust indicator of global cluster
properties and has a different dependence on cosmological parameters than the
central decrement. However, the total decrement has been difficult to measure
to date.  Single dish experiments can, in principle, measure the total
decrement, but in practice, the need to subtract large backgrounds have
resulted in limited signal to noise (Mason et al. 2001; Myers et al. 1997). The most precise SZE
observations have so far been obtained with interferometers.  Interferometric
(Joy et al. 2001; Kneissl et al. 2001; Grego et al. 2000; Reese et al. 2000;
Carlstrom et al. 2000; Jones et al. 2001; Birkinshaw 1998 and references
therein) observations offer several advantages over observations made with
single dishes, in terms of better control over systematics and contamination
effects. However, at present, the angular size that can be imaged is small,
relative to the angular size of the massive clusters that can be studied at
the
current sensitivities. Hence, these observations typically do not cover the
whole cluster, and instead measure only the central decrement.

The prospects for measuring the total SZ flux decrement will dramatically
improve in the next few years.  Forthcoming SZE (e.g., DASI, CBI) and
cosmic microwave background (CMB) experiments will not only measure the total
SZD of an extended cluster sample, but, in many cases, will also resolve the
temperature fluctuation over the angular size of the whole cluster.

 For example MAP (\cite{MAP}) will detect tens of  clusters and
resolve a few of them; Planck (Bersanelli
et al. 1996; Mandolesi et al. 1998) will detect $\sim 10^{4}$ clusters and
resolve and measure the total SZD of $1$\% of them (\cite{SZPlanck01}).  
An experiment such as the proposed
Center for study of Cosmic structure (CSCS; Page at al. 2001) could map
the CMB anisotropy over 100deg$^2$ with a resolution of 1.7'. This
would allow {\it all} galaxy clusters of masses above $4\times 10^{14}$ M$_{\odot}$
in the CMB map region to be detected through the SZ effect. The experiment would
also determine spectroscopic redshift of more that 300 clusters, and
implement X-ray follow up to determine their X-ray temperatures.

In this paper, we investigate the scientific potential of measuring
the cluster total SZ flux decrement in the context of a well studied
cluster sample for which X-ray and optical follow-up are available.
As illustrated in \S~3, the total SZ flux decrement ($\Delta S_{\nu}$)
is an integrated quantity over the whole cluster, not just along the
central line of sight, and should thus be more sensitive to global
cluster properties. In particular, $\Delta S_{\nu}$ probes a unique
combination of the physical parameters of the hot intra--cluster gas,
different from those inferred from X--ray observations, and from other
data.  This allows the construction of new scaling relations.  In this
paper, we examine these scaling relations in detail.  We are motivated
by the forthcoming sample of clusters with measured SZ decrements, as
well as by the strong need for observational tests of the assumption
that clusters can be used as ``standard candles''.  Our goals are (1)
to predict scaling relations that involve the SZD, together with their
scatter, and (2) to study how these scaling relations depend on
assumptions about the cluster structure (i.e. that clusters are
standard candles), as well as on the underlying cosmological
parameters.  In particular, we quantify these dependencies in the
context of a well studied cluster sample, such as those expected to be
available in the near future, for which it will be possible to combine
SZE and X--ray observations.

This paper is organized as follows. In \S~\ref{sec:model}, we describe
our cluster model and its input model parameters.  In
\S~\ref{sec:observables}, we summarize the main observables that the model
predicts: redshift ($z$), virial temperature ($T$), Sunyaev--Zeldovich
decrement ($\Delta S_\nu$), and angular size ($\theta$).  In
\S~\ref{sec:fp}, we show that in this model at a fixed $z$ in
($T,\Delta S_{\nu},\theta$) space, clusters are expected to be
distributed on a ``fundamental plane'' with nearly negligible scatter
in the $\Delta S_\nu-T$ projection, and that the plane's orientation
depends on redshift and on the background cosmology.  In
\S~\ref{sec:clusters}, we parameterize deviations from our cluster
model caused by additional energy input, or by a lack of full
virialization, and quantify their effects on the scaling relations.
Similarly, in \S~\ref{sec:cosmology}, we quantify the effects on the
scaling relations caused by varying several cosmological parameters.
In \S~\ref{sec:discussion}, we discuss our results and the
implications of this work. Finally, in \S~\ref{sec:conclusions}, we
summarize our conclusions.

\section{Model Ingredients}
\label{sec:model}

\subsection{Spherical Collapse}
\label{sec:tophats}

Generally, galaxy clusters models are based on the collapse of a
spherical top--hat perturbation (e.g., \cite{P80}).  In this model,
the average density $\bar\rho_{\rm vir}$ enclosed within the radius
$R_{\rm vir}$ of a cluster of total (dark matter + baryon) mass
$M_{\rm vir}$ that forms at redshift $z_f$ is given by (e.g., Kitayama
\& Suto (1996) hereafter KS96)
\be
\frac{\bar\rho_{\rm vir}}{\bar\rho(z_f)}=
18\pi^2 \wp(z_f),
\label{eq:ks96}
\ee
where the factor $\wp(z_f)$ describes departures from a standard
Einstein de-Sitter universe, and is given by
\begin{equation}
\label{eq:bzf}
\wp(z_f)\approx 1+0.4093 W_f^{0.9052}, 
\end{equation}
where
\begin{equation}
 W_f  =1/\Omega_f-1,
\end{equation}
\begin{equation}
\Omega_f=\frac{\Omega_0(1+z_f)^3}{\Omega_0(1+z_f)^3\!+\!(1-\Omega_0-\Omega_\Lambda)(1+z_f)^2\!+\!\Omega_\Lambda}.
\label{eq:omegaz}
\end{equation}
In the above equations, $\bar\rho(z_f)=\Omega_0\rho_{\rm
crit,0}(1+z_f)^3$ is the mean matter density at redshift $z_f$, and
$\Omega_0$ and $\Omega_\Lambda$ are the present day densities of
matter and cosmological constant, in units of the critical density
$\rho_{\rm crit,0}$.

We assume that the cluster is isothermal, and that the gas acquires the 
virial temperature of the halo.  In this case, the temperature is related
to the mass and virial radius by the virial theorem,
\be
\left(\frac{T_{vir}}{\rm 1~keV}\right)
\left(\frac{R_{vir}}{\rm 1~Mpc}\right)=
0.88 
\left(\frac{M_{\rm vir}}{\rm 10^{14}~M_\odot}\right)\;,
\label{eq:tvir}
\ee
where we have assumed that the gas has a mean molecular weight of
$\mu=0.59$, appropriate for a fully ionized mixture of H and He with a
number density ratio of $n_{\rm He}/n_{\rm H}=0.08$.  Combining
equations (\ref{eq:ks96})--(\ref{eq:tvir}), the temperature of a
cluster is given in terms of its mass and formation redshift,
\begin{eqnarray}
\nonumber
\frac{T_{\rm vir}}{[KeV]}&=&
0.62(1+z_f)\left[\frac{\Omega_0h^2}{0.17}\right]^{1/3}\\
&\times&[\wp(z_f)]^{1/3}\left(\frac{M_{\rm vir}}{10^{14}{\rm M_{\odot}}}\right)^{2/3}[{\rm KeV}]
\label{eq:TzfM}
\end{eqnarray}
Here $h\equiv H_0/100~{\rm km~s^{-1}~Mpc^{-1}}$ is the Hubble
constant.  This equation quantifies the intuition that clusters
virializing at an earlier epoch are hotter than clusters of the same
mass that virialize at later times. Following KS96, we note that, in
fact, clusters of a given mass $M$ observed at a redshift $z_o$ form
over a range of earlier redshifts $z_f\geq z_o$.  While equations
(\ref{eq:tvir}-\ref{eq:TzfM}) give the temperature and mass of a
cluster when it first forms and virializes, the cluster is
subsequently expected to evolve.  In principle, both the mass and the
temperature can change between $z_f$ and $z_o$.  As argued in KS96,
numerical simulations (e.g., \cite{Evr90,Navarro95})
indicate that the temperature does not significantly evolve after
virialization.  We therefore make the assumption here that the final
temperature of the cluster at redshift $z_o$ is equal to $T_{\rm
vir}$.  Below, we will consider models that describe departures from
this assumption (see \S~2.3).

However, clusters do grow in mass after virialization.  To account for
this growth, we assume that the ratio between the final mass $M$ and
the mass at virialization $M_{\rm vir}$ is a ``universal'' constant
$M_{\rm vir}=f_M\times M$ for all clusters (see Lacey \& Cole 1993; we
will relax this assumption below).  Following Viana \& Liddle (1996),
we set $f_M=0.75$, which results in the best fit to the
mass--temperature relation (both for its slope and scatter) in cluster
simulations.

\subsection{Mass Function and Age Distribution}
\label{sec:massfunction}

The mass and formation redshift of a cluster are not directly
measurable quantities.  Nevertheless, it is possible to compute the
cluster mass function, $dN/dM(M,z_o)$ (e.g.,
\cite{PS74,ShethMoTormen99,Jenkinsetal00}), and also the statistical
distribution of the formation redshifts $z_f$ of clusters of a given
mass $M$ observed at redshift $z_o$, $dN/dz_f(M,z_o)$ (e.g.,
\cite{LC93,Sasaki94}).  In what follows, we will use the standard
Press \& Schechter (Press \& Schechter 1974; PS) mass function, with
fitting formulas for the cosmological transfer function (Sugiyama
1995), and for the critical overdensity $\delta_c(z)$ for collapse
(KS96):
\begin{eqnarray}
\nonumber
\delta_c(z)&=&\delta_0(z)(1+z)g(0)/g(z)\\
\delta_0&\simeq&\frac{3(12\pi)^{2/3}}{20}(1+0.0123 \log_{10}\Omega_f).
\label{eq:deltac}
\end{eqnarray}
Here $(1+z)/g(z)$ is the usual linear theory growth factor
(\cite{P80}).  To obtain the formation redshift distribution for
clusters of mass $M$ at redshift $z_o$, we follow equation 2.26 in
Lacey \& Cole (1993; hereafter LC93), and define ``formation
redshift'' as the redshift at which clusters first acquire a fixed
fraction $f_M=0.75$ of their final mass.  

Our main motivation for the above choices is ``technical'': the
convenient semi--analytical derivation of $dN/dz_f$ is only applicable
to the mass function in the standard Press--Schechter theory; at
present no analogous derivation exists for the improved mass
functions, such as that of Jenkins et al. (2000).  We also note that
our qualitative conclusions below do not crucially depend on the shape
of the distributions $dN/dz_f$ and $dN/dM$.  Nevertheless, it is
important to emphasize that improvements over the PS mass function
have already been made, and simulations have possibly uncovered
differences in the abundance of massive clusters (e.g., Seth, Mo \&
Tormen 1999, \cite{Jenkinsetal00}).  Our analysis can be
straightforwardly generalized to different choices for the mass
function and formation redshift distribution, as the latter becomes
available.

\subsection{Deviations from  Simple Spherical Collapse}
\label{sec:clusterphysics}

The simple model described above assumes that clusters are fully
virialized objects whose abundance and structure is dictated by
gravitational physics alone.  Lack of full virialization, on--going
mergers, or feedback from galaxy formation can modify these
predictions.  For example, the above simple scaling relations,
together with the assumption that the X--ray luminosity $L_X$ is
dominated by Bremsstrahlung emission, predicts $L_X \propto T^2$,
while observations indicate the steeper relation $L_X \propto T^3$
(see, e.g., Kaiser 1991; Evrard \& Henry 1991; \cite{BN98}).  Although different explanations are still
possible (\cite{B01}) this steep slope suggests additional heat input
to the gas prior to virialization, which preferentially lowers the
central density, and therefore the X--ray emissivity, in smaller
clusters (see, e.g., \cite{K91}; \cite{BEM01} and references therein
for recent work on this subject).

To model deviations from our simplified model, we here introduce two
additional parameters, $\xi$ and $\alpha$, by generalizing equation
(\ref{eq:TzfM}) as follows:
\be
T_{\rm vir} \propto (1+z_f)^\alpha M_{\rm vir}^{1/\xi}.
\label{eq:TzfM2}
\ee
The choice of $\xi=1.5$ and $\alpha=1$ corresponds to the original
equation~(\ref{eq:TzfM}), while different values of $\xi$ and $\alpha$
describe deviations from the simplest model.  By design, both of these
parameters capture departures from purely gravitational virial
equilibrium (cf. eq.~\ref{eq:tvir}).  The first parameter, $\xi$,
mimics the effect of heat input.  Preferential heating of small
clusters corresponds to flattening the $T-M$ relation, i.e. to values
of $\xi\geq 1.5$.  This type of flattening has already been observed,
e.g., $\xi \sim 1.6$ (\cite{Mohrevrard97}), $\xi=1.72$
(\cite{MTKPC01}; from analysis of cluster simulations), $\xi\simeq
1.81$ (\cite{XuJinWu01}), and $\xi=1.98\pm 0.18$ (\cite{MME99}); 7\%,
10\%, 20\%, and 25\% variations respectively.  Within the context of
the spherical top hat model, this discrepancy is partly alleviated by
the fact that smaller clusters form over a broader range of redshifts,
and are on average hotter than larger cluster (which, on average, form
closer to $z=z_o$). We find that the effective slope $\xi_{\rm eff}$,
under the assumption of $f_M=0.75$, would be $\xi_{\rm eff}\sim
1.6$. Nevertheless it is important to model possibly still larger
deviations from the $\xi=1.5$ scaling behavior.

The second parameter, $\alpha$, quantifies deviations from the
assumption that clusters at all redshifts are fully virialized. It is
reasonable to consider, for example, the possibility that clusters at
higher redshifts are more likely to be detected in the process of
their initial collapse/assembly, and are less likely to be fully
virialized than their lower--redshift counterparts.  This effect is
mimicked by a choice of $\alpha\leq 1$, implying cluster are
``colder'' at higher redshifts.  Interestingly, there are examples of
clusters hotter than one would expect if they were virialized (and can
be taken as evidence for on--going mergers, or some other form of
energy injection; e.g., \cite{Tucker98,Roettiger97,Evr96}); this case
is mimicked by a choice of $\alpha\geq 1$.

\subsection{Modeling the Scatter}
\label{sec:scatter}

The model described above is ``deterministic'': the only source of
scatter it predicts in the observables is that caused by the fact that
clusters have a distribution of different formation redshifts.  This
scatter is a direct consequence of cosmological initial conditions
(which translates into a ``scatter'' by subsequent gravitational
collapse), and probes the high--$\sigma$ tail of the primordial power
spectrum (\cite{VKMB}).  For example, in the mass--temperature or
size--temperature relations, this causes a scatter whose magnitude
monotonically increases with decreasing temperature.

It appears that this scatter alone is sufficient to account for the
scatter in the observed $M-T$ relation
(\cite{Vianaliddle96,Horneretal98,FRB00,XuJinWu01}), and, for a
suitable choice of $\sigma_8$, for the observed scatter in the
size--temperature relation (\cite{VKMB,Mohrevrard97,M00}).
Nevertheless, it is interesting to contrast the predictions of this
``cosmological'' scatter with scatter that can be caused by different
physics.  For example, in a recent analysis of hydrodynamical cluster
simulations (\cite{Mathiesen00}), the X--ray temperatures do not
appear to depend on the formation redshift (although see Evrard 1990
and Navarro et al. 1995, who find that cluster temperatures do not
change after the cluster forms).  These simulations are still
inconclusive, since the expected correlation between formation
redshift and cluster temperature is smaller than the scatter due to
small--number statistics in the simulations.  Furthermore, the
observational scatter might even be due mostly to uncertainties in the
mass, temperature (or size) determinations. We here nevertheless
consider alternative origins for the scatter as a possibility.

We will distinguish two extreme cases as a {\it deterministic}
scenario, in which the cluster temperature depends on the formation
redshift, with no additional source of scatter; and a {\it stochastic}
scenario, in which the temperature scales directly with the observed
redshift, and there is some additional source of scatter.  In the
latter case, the additional scatter may simply be observational, or it
may be caused by processes that influence cluster temperatures, such
as feedback from galaxy formation, other forms of heating or cooling,
or simply that $f_M$ is not constant from cluster to cluster.
Regardless of the source of this stochasticity, we assume that on
average, the X--ray temperature of a cluster depends on its mass,
redshift of observation and cosmology through (\ref{eq:TzfM}), with
the substitution $z_f\longrightarrow z_o$, but that there is, in
addition, an intrinsic scatter around this mean $T(z_o)$ relation. We
model this scatter as random deviations in the temperature from
cluster to cluster, where the deviations $\Delta T$ are Gaussian
distributed with a fractional $r.m.s.$ width $x\equiv\Delta T/T$.

In reality, the scatter seen in cluster scaling relations is likely
due to a combination of a deterministic effect (such as in our {\it
deterministic} scenario ) and a random variation from cluster to
cluster (such as in the {\it stochastic} scenario). For example, as we
will discuss in \S~5.1 below, approximately the same scatter in the
$M-T$ distribution could be obtained within the {\it deterministic}
scenario (with $f_M=0.75$; $x=0$), or from a different scenario in
which the formation redshift distribution was narrower
(i.e. $f_M>0.75$), but the cluster temperature had some additional
random variation (i.e. $x>0$).  The parameter $x$, together with
$f_M$, quantifies how much of the observed scatter is due to
cosmological initial conditions {\it vs.} other sources.  The {\it
deterministic} scenario corresponds to ($f_M=0.75;x=0$); the {\it
stochastic} scenario corresponds to ($f_M=1;x>0$); and in-between
cases are described by ($f_M<1;x>0$).

\subsection{Summary of Model Parameters}
\label{sec:parameters}

In principle, one would like to set all parameters of our model to be
free, and investigate what joint constraint can be imposed from
observations.  However, the number of parameters makes this approach
impractical, and we are forced to impose constraints using other
observations.  We assume the background cosmology to be a flat
$\Lambda$ cold dark matter (CDM) model (as supported by recent CMB data;
\cite{boomerang1,boomerang3,boomerang4}), described by the parameters
$\Omega_0$ and $h$.  Unless otherwise stated, we will assume that the
combination $\Omega_0 h^2$ is a constant (justified by the fact that
forthcoming CMB experiments will constrain this combination to better
than $5\%$).  We also assume a baryon fraction $\Omega_b h^2=0.02$,
consistent with recent D/H measurements (e.g., \cite{BT98}).  This
leaves $h$ as the only free cosmological parameter, and we will
further impose $0.2<h<0.9$. The primordial power spectrum is specified
by the normalization $\sigma_8$ and slope $n$; however, we here assume
$n=1$ (supported by recent CMB data), and, unless otherwise stated, we
also adopt $\sigma_8=0.495 \Omega_0^{-0.60}$ (inferred from local
cluster abundance, \cite{Vianaliddle99,Pierpaoli00}).

In summary, our model is fully specified by the following 7
parameters:
\begin{itemize}
\item The cosmological parameters $\Omega_0,h,$ and $\sigma_8$; two of which
can be eliminated by other constraints.
\item The parameters $f_M$ and $x$ that describe scatter in the scaling
relations attributable to cosmological initial conditions, and to
other sources, respectively
\item The parameters $\xi$ and $\alpha$ that describe departures from
the fully virialized simple top--hat collapse models.
\end{itemize}

For illustrative purposes we define a ``fiducial model'' by the
following choice of parameters: $h=0.65$, $\Omega_0h^2=0.17$,
$\Omega_0=0.4$, $\sigma_8=0.86$, $f_M=0.75$, $x=0$, $\xi=1.5$, and
$\alpha=1$ for the {\it deterministic} model ; for the {\it
stochastic} model we chose $x=0.13$ and $f_M=1$ and everything else
remains unchanged.  In addition to these parameters, to predict the
Sunyaev--Zeldovich decrement (see \S~\ref{sec:observables} below), we
need to know the mass fraction of baryons that constitute the hot gas
in the intra--cluster medium (ICM).  This fraction ($f_{ICM}$) can be
inferred from a comparison of the clusters X--ray flux and X-ray
temperature; here we assume $f_{ICM}\sim 0.2$ with a weak temperature
dependence: $f_{ICM}=0.2 T^{0.266}(h/0.5)^{-3/2}$ (\cite{MME99}).  We
note here that the value of $f_{ICM}$ should be a lower limit to the
ratio $\Omega_b/\Omega_0$; the temperature dependence reflects the
fact that low mass clusters are more likely to loose their gas from
energy-injection or pre-heating mechanisms, while the dependence on
$h$ depends on the method employed to measure it. The method of Mohr
et al. (1999) is a determination for the whole clusters, while the
method of e.g., Grego et al. (2000b) yields a different $h$ dependence
but is based on the central SZ decrement.  We have checked that the
results we present here do not depend qualitatively on the specific
choice of the dependence of $f_{ICM}$ on $h$.

\section{Observables}
\label{sec:observables}

The observables of a cluster that we utilize in this paper are the
redshift ($z_o$), temperature ($T$), Sunyaev--Zeldovich total flux
decrement ($\Delta S_\nu$), and angular size ($\theta$).  Within our
model, the redshift and temperature are independent parameters and
have already been specified.  Cluster redshifts can be obtained
observationally e.g., from their member galaxies out to $z_o\sim 1$
through deep optical and near infrared observations, and cluster
temperatures can be obtained from X--ray spectra out to similar
redshifts.  Here we define the Sunyaev--Zeldovich decrement (SZD) and
the angular size.

The SZD along a line of sight to a cluster in the Rayleigh-Jeans
regime of the CMB is given by (e.g., \cite{Holderetal00})
\be
\frac{\Delta T_{\rm CMB}}{T_{{\rm CMB}}}=\frac{g k_B \sigma_T}{m_e c^2}\int dl n_e(l)T_e(l),
\label{eq:deltat}
\ee
where $g=-2$, $m_e$ is the electron rest mass, $c$ is the speed of
light, $\sigma_T$ is the Thomson cross section, $T_{\rm CMB}$ is the
CMB temperature, $T_e$ is the electron temperature, $n_e$ is the
electron number density, and the integral is along the line-of-sight.
Expressed in terms of a decrease in the observed flux, $\Delta S_\nu
=2 k_B \Delta T_{\rm CMB} \nu^2/c^2 d\Omega$ (where $d\Omega$ is the
smaller of the solid angle of the observations and the solid angle
subtended by the cluster), equation~(\ref{eq:deltat}) implies that the
total observed SZ flux decrement ($\Delta S_{\nu}$) for a cluster can
be related to its total mass via
\be
\Delta S_\nu=\left(
\frac{g 2 k_B^2 \nu^2 \sigma_T T_{{\rm CMB}}}{\mu_e m_e m_p c^4}
\right)
f_{\rm ICM} \frac{T_e M}{d_A(z)^2},
\label{eq:Sth}
\ee
where $\nu$ is the frequency, $\mu_e=1.15$ is the mean molecular
weight per electron, $m_p$ the proton rest mass, $d_A$ is the angular
diameter distance, $z$ is the cluster redshift, $f_{\rm ICM}$ is the
ICM mass fraction and $M$ denotes the mass enclosed within the virial
radius of the halo. Equation~(\ref{eq:Sth}) also assumes that the
electron density weighted mean temperature $T_e$ can be identified
with the virial temperature as given in equation~(\ref{eq:tvir}) (this
is satisfied by definition in our assumed isothermal clusters; we
implicitly assume that $T_e$ also equals the actual observed $X$-ray
temperature $T_X$).  Finally equation~(\ref{eq:Sth}) assumes that the
solid angle of the observations is larger than, or equal to the solid
angle subtended by the cluster.  Using equation~(\ref{eq:tvir}),
equation~(\ref{eq:Sth}) can be rewritten in convenient units in terms
of the observables $z$ and $T_X$ as:
\begin{eqnarray}
\frac{\Delta S_{\nu}}{[{\rm \mu Jy}]}&=&0.26(1+z_o) \left(\frac{\nu}
{[{\rm GHz}]}\right)^2\\ \nonumber	
&\times& f_{\rm ICM} \frac{M}{[10^{14}{\rm M_\odot}]}\frac{T_X}{[{\rm KeV}]}\frac{h^2}{d_A'^2}
\label{eq:convevientunits}
\end{eqnarray}
where we have introduced the dimensionless angular diameter distance
$d_A'=d_A H_0/c$, and $z_o$ is the cluster redshift. The observable
quantities here are $z_o$, $T_X$, and $\Delta S_{\nu}$.

In our model, the angular size of a cluster simply corresponds to the
virial radius, $\theta=R_{\rm vir}/d_A$. The spherically averaged
profiles (of, e.g., temperature) found in numerical simulations do
indicate the presence of a virialization shock, as expected from the
top--hat collapse model.  However, the shock tends to be weaker, and
located at larger radii (see, e.g., Bryan \& Norman 1998).
Nevertheless, we here assume that the angular size is given by the
virial radius, as obtained from the top hat collapse model (see
further discussion below).

\begin{equation}
\label{eq:theta}
\frac{\theta}{[ {\rm deg} ]} 
= \frac{\frac{\Delta S_{\nu}}{[ \mu {\rm Jy}]} f_M d_A'}{f_{\rm ICM} 15.6 
(1+z_o) (\frac{\nu}{[{\rm GHz}]})^2 (\frac{T_X}{[{\rm KeV}]})^2 h}
\end{equation}

Consistent with our earlier assumption, we have assumed that the
cluster grows in mass by a factor of $1/f_M=1.3$ between $z_f$ and
$z_o$.
In practice, this angular size can be measured accurately only if a
cluster is resolved, and therefore an estimate of $\theta$ will be
available only for a fraction of the clusters in a dataset. For
example, the Planck Surveyor satellite will resolve only about
$\sim1\%$ of all SZE clusters (\cite{SZPlanck01}), but the CSCS, with
a 1.7' resolution, will resolve {\it most} clusters at $z\lap 1$
(\cite{ACT}).  Angular sizes can, in principle, also be obtained from
clusters resolved in X--ray observations. In practice, this involves
measuring the profile and fitting to a model; this fit will be
constrained essentially by the core radius, rather than the virial
radius (although the latter can, of course, be inferred by
extrapolation once the radial profile was fit to the model).

It is likely that in a given sample, not all observables will be known
for each cluster, and the model prediction for some observables (e.g.,
for the angular size) are less robust than for others.  Below, we will
briefly consider the ideal case, in which all four observables
($z,T,\Delta S_\nu,\theta$) are reliably measured for a cluster
sample.  We will then concentrate on cases where only the subsets
($z,T,\Delta S_\nu$) or ($z,\Delta S_\nu,\theta$) have been measured.
The first of these is relevant e.g., to a sample of clusters detected
in the interferometric SZE survey of Holder et al. (2000), or to the
MAP/Planck/CSCS cluster sample with X-ray follow-up and reliable
temperatures; the second case is relevant to a sample of well resolved
SZE clusters (e.g., those of the CSCS) that are too distant ($z\gsim
1$) to obtain e.g., reliable temperatures.  Current cluster samples
exist where a combination of cluster sizes and temperatures are
measured, and have been analyzed in the literature (see, e.g.,
\cite{Mohrevrard97,VKMB}).

\myputfigure{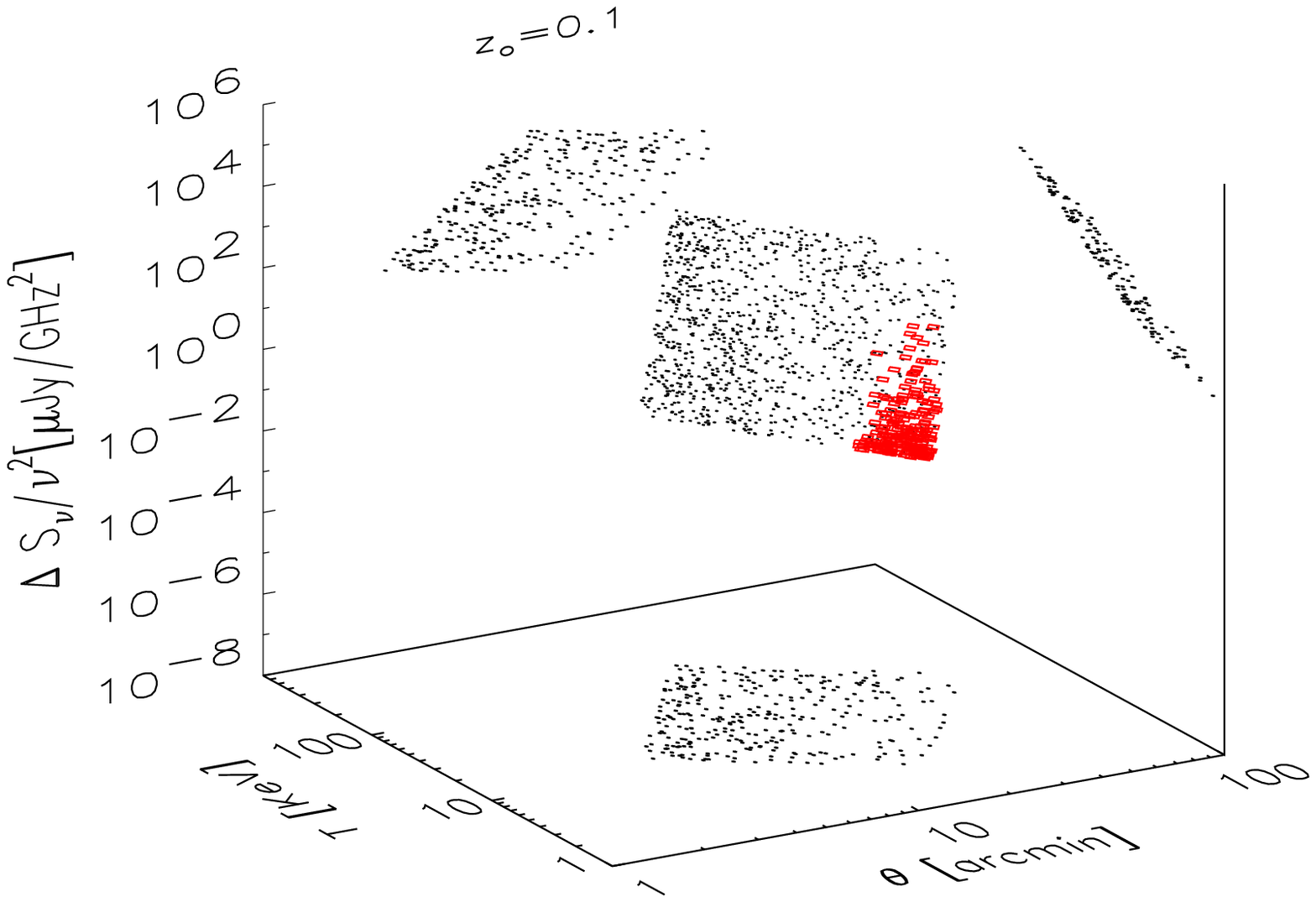}{3.2}{0.50}{-25}{-00}
\figcaption{\label{fig:hyperplane}
If clusters are standard candles, they populate a hypersurface in the
4 dimensional space of the observables ($T_X$, $\Delta S_\nu$,
$\theta$ and $z$). Here, we show a slice of this space at the constant
redshift $z_o=0.1$.  In this slice, clusters are constrained to be on
a ``fundamental plane''.  The figure also shows the projection of this
plane onto the ``walls'' of the box, defining the three scaling
relations ($\Delta S_\nu - T_X$), ($\theta-\Delta S_\nu$), and
($\theta - T_X$).  Not all locations in the fundamental plane are
equally likely to be occupied: the filled symbols show the
distribution of 200 Monte-Carlo generated clusters for our fiducial
model. Deviations from the top--hat collapse model, or different
choices of the cosmological parameters, cause measurable deformations
of the FP such as a shift in its position and orientation, and can
also introduce a random scatter around it.}

\section{The Fundamental Plane of Galaxy Clusters}
\label{sec:fp}

\subsection{The Fundamental Plane in the Fiducial Model}
\label{sec:fp0}

At any given redshift $z_o$, the equations~(6), (11) and (12) (i.e. in
the {\it deterministic} scenario) express the three observables $T_X$,
$\Delta S_\nu$ and $\theta$ in terms of the two variables $z_f$ and
$M$.  For a given frequency channel $\nu$ of the SZE survey, we assume
that the ICM mass fraction has the functional form described in
\S~2.5 and that the cosmological parameters ($\Omega_0,h,$ and
$\sigma_8$) and $f_M$ are known.  These relations -- three equations
between the five variables ($T_X$, $\Delta S_\nu$, $\theta$, $M$, and
$z_f$) -- allow a study of the distribution of clusters in several
useful projections of this 5--dimensional parameter space.  In
particular, clusters define a 3--dimensional manifold in the
4--dimensional subspace of the four observables ($T_X$, $\Delta
S_\nu$, $\theta$, $z$).  It is, however, more practical to consider
the three dimensional subspace ($T_X$, $\Delta S_\nu$, $\theta$), and
regard the cluster redshift as an evolutionary parameter.  In this
case, for a given redshift $z_o$, clusters can only populate a
half--plane in ($T_X$,$\Delta S_\nu$, $\theta$)--space.  The
restriction to a half--plane arises from the requirement that clusters
form prior to the redshift at which they are observed, $z_f\geq z_o$.

In order to visualize this half--plane, which we will hereafter refer
to as the ``fundamental plane'' (FP), in Figure~\ref{fig:hyperplane}
we show an example of the distribution of clusters observed at the
fixed redshift $z_o=0.1$.  The figure also shows the projection of the
FP onto the ``walls'' of the box, which define the three scaling
relations ($\Delta S_\nu - T_X$), ($\theta-\Delta S_\nu$), and
($\theta - T_X$). Of course, not all locations in the FP are equally
likely to be occupied: cluster temperatures, virial radii and SZ
decrements cannot assume any arbitrary value between $-\infty$ and
$+\infty$; smaller clusters are more numerous than high temperature
ones; and the width of the probability distribution of $z_f$ decreases
monotonically for larger clusters. As a result, clusters preferably
populate a portion of the FP. In Figure~\ref{fig:hyperplane}, the
filled symbols show the distribution of 200 clusters, selected
randomly at $z_o=0.1$ using a Monte--Carlo method.

If all observable quantities $T_X$, $\Delta S_\nu$, $\theta$ and $z$
are measured in a cluster sample, and the top hat collapse model is
the correct description of the physical properties of clusters, then
the clusters define the FP shown in Figure~\ref{fig:hyperplane},
determined solely by the cosmological parameters. Deviations from the
top--hat collapse model, or different choices of the cosmological
parameters, can cause a shift in the position and orientation of the
FP, and can also introduce a random scatter around it.  We will argue
that it is possible to quantify such deviations from the simple
top--hat model, and gain insight into both cluster physics and
cosmological parameters, by studying the departures of the FP from its
position and orientation in our ``fiducial'' model.  As an example,
$\alpha < 1$ would modify the dependence of the FP on redshift $z$;
while $x > 0$ would induce a scatter around the FP.  In the rest of
this paper, we will quantify these statements, by concentrating on
different projections of the FP.

\subsection{The Fundamental Plane and Cluster Scaling Relations}
\label{sec:fpprojections}

As mentioned above, the projections of the FP define three scaling
relations, ($\Delta S_\nu - T_X$), ($\theta-\Delta S_\nu$), and
($\theta - T_X$).  The last of these relations can be analyzed using
existing cluster samples, and have been studied by various authors
(e.g., \cite{Mohrevrard97,VKMB}).  In this paper, we focus on the
usefulness of the first two scaling relations, which will be
observationally available from future SZE cluster surveys.

\myputfigure{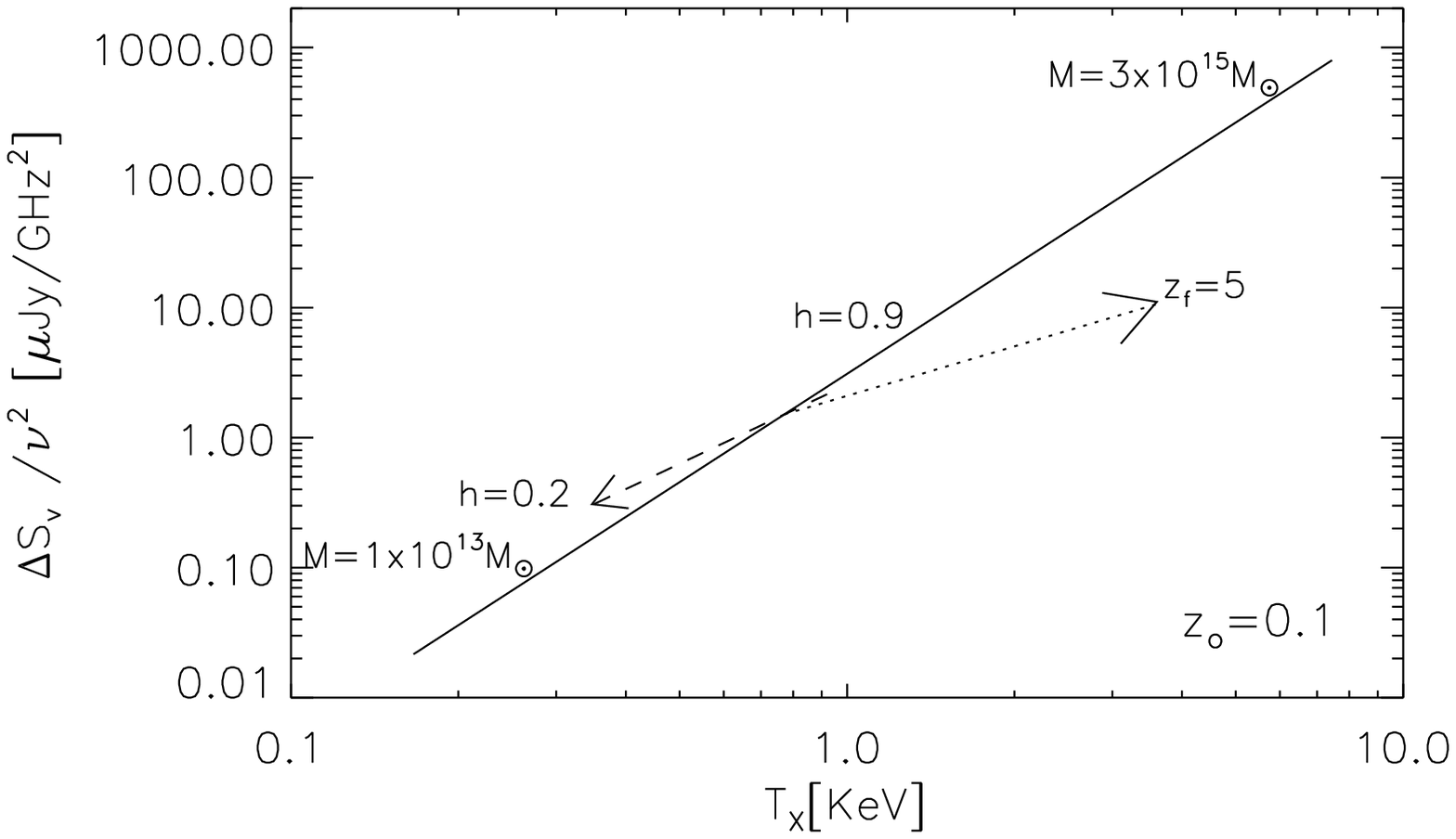}{3.2}{0.50}{-25}{-00}
\figcaption{\label{fig:SZthall}
Theoretical expectations for the projection of the fundamental plane
defining the ($\Delta S_\nu - T_X$) scaling relation. The solid line
shows the location of clusters in our fiducial model at redshift
$z_o=0.1$ with masses between $10^{13}~{\rm M_\odot}\leq M\leq 3\times
10^{15}~{\rm M_\odot}$, assuming that they formed at the redshift at
which they are observed, $z_f=z_o=0.1$.  The dotted arrow shows how
the location of a $2\times 10^{14}~{\rm M_\odot}$ cluster depends on
its formation redshift for $0.1<z_f<5$.  Similarly, the dashed arrow
shows how the location of this cluster would change with the Hubble
constant for $0.2<h<0.9$.  Changes in $\Omega_0$ or $\sigma_8$ mainly
affect how the clusters are distributed in the direction of the dotted
arrow.  Note that this arrow is nearly parallel to the solid line;
this suggests that the scatter in the ($\Delta S_\nu - T_X$) should be
small.}

\myputfigure{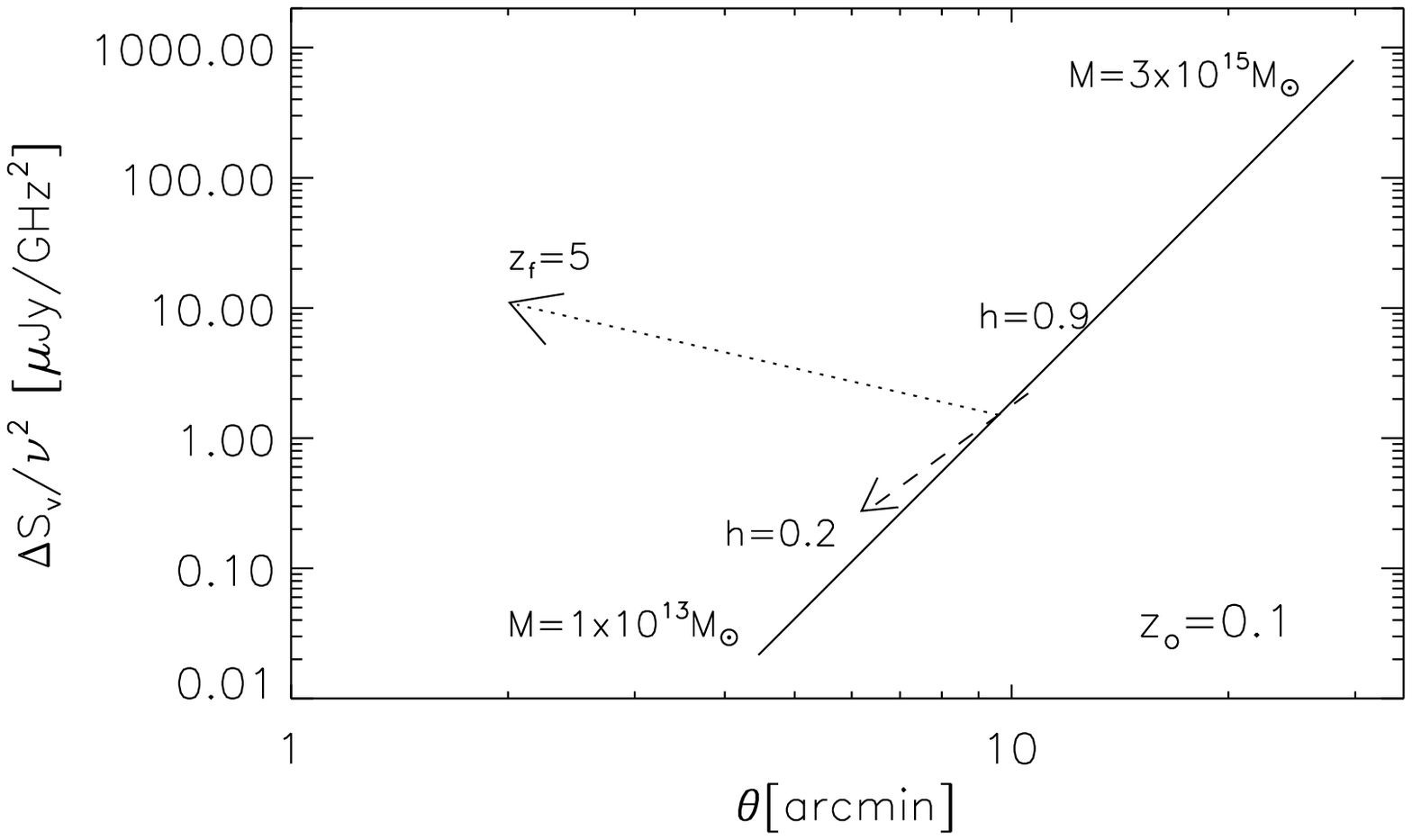}{3.2}{0.50}{-25}{-00}
\figcaption{\label{fig:SZ-theta}
Theoretical expectations for the projection of the fundamental plane
defining the ($\Delta S_\nu - \theta$) scaling relation.  The solid
line, and the dotted and dashed arrows describe the same clusters as
in Figure~\ref{fig:SZthall}. Note that in difference from
Figure~\ref{fig:SZthall}, the dotted arrow, representing increasing
formation redshifts, is nearly orthogonal to the solid line. This
suggests that the scatter in the ($\Delta S_\nu - \theta$) should be
large, and sensitive to $\Omega_0$, $\sigma_8$ and primordial
non-gaussianity (the same conclusion holds for the size--temperature
relation, as can be seen from Figure~\ref{fig:hyperplane}; see also
\cite{VKMB}).}
\vspace{\baselineskip} 

Figures~\ref{fig:SZthall} and \ref{fig:SZ-theta} show the theoretical
expectations for the two projections of the FP that involve the
Sunyaev--Zeldovich decrement, together with the expected position of
clusters in our fiducial model.  Figure~\ref{fig:SZthall} shows the
projection of the FP onto the ($\Delta S_\nu - T_X$) plane. The solid
line shows the location of clusters at redshift $z_o=0.1$ with masses
between $10^{13}~{\rm M_\odot}\leq M\leq 3\times 10^{15}~{\rm
M_\odot}$ in this plane, assuming that these clusters formed at the
redshift at which they are observed, $z_f=z_o=0.1$.  The dotted arrow
shows how the location of a $2\times 10^{14}~{\rm M_\odot}$ cluster in
this plane depends on its formation redshift for $0.1<z_f<5$.  As the
arrow demonstrates, both the temperature and SZD are increased for
clusters that form earlier.  Similarly, the dashed arrow shows how the
location of this cluster would change with the Hubble constant for
$0.2<h<0.9$.  Both the temperature and SZD decrease if the Hubble
constant is lower.  Note that changes in $\Omega_0$ and $\sigma_8$
mainly affect how clusters are distributed in formation redshift,
i.e. along the direction of the dotted arrow. Since this is nearly
parallel to the solid line, a change in $\Omega_0$ or $\sigma_8$ is
nearly degenerate with a change in cluster mass.  This suggest that
the expected scatter in the ($\Delta S_\nu - T_X$) relation should be
small, and insensitive to $\Omega_0$ and $\sigma_8$. Notice also that
low values of $h$ make clusters appear colder, mimicking an unphysical
formation redshift $z_f<z_o$.

Figure~\ref{fig:SZ-theta} shows the projection of the FP onto the
($\Delta S_\nu - \theta$) plane.  As in Figure~\ref{fig:SZthall}, the
solid line shows the location of clusters at redshift $z_o=0.1$ with
masses between $10^{13}~{\rm M_\odot}\leq M\leq 3\times 10^{15}~{\rm
M_\odot}$, while the dotted and dashed arrows demonstrate the
dependence of the location of a $2\times 10^{14}~{\rm M_\odot}$
cluster on its formation redshift and on the Hubble constant.  Note
that in difference from Figure~\ref{fig:SZthall}, the dotted arrow,
representing increasing formation redshifts, is nearly orthogonal to
the solid line. This suggests that the scatter in the ($\Delta S_\nu -
\theta$) should be large, and sensitive to $\Omega_0$, $\sigma_8$ and
primordial non-gaussianity. The same conclusion holds for the
size--temperature relation, as can be seen from
Figure~\ref{fig:hyperplane} (see also \cite{VKMB}).

\section{Probing the Internal Physics of Clusters}
\label{sec:clusters}

The idealized model for clusters described in \S~\ref{sec:tophats} is
based on the collapse of a spherical top hat perturbation, and is the
simplest model to relate cluster observables.  It is important to have
observational tests for departures from this model; both in order to
understand the structure of clusters themselves, and also to allow
clusters to be used as ``standard candles'' in cosmological studies.
As discussed above, all four observables ($\Delta S_\nu,T,z,\theta$)
might not be available for every cluster.  This prompts us to consider
what constraints are possible both with and without knowing the
angular size $\theta$.  In this section we will assume that
cosmological parameters are known and use the deformations of the FP
to test whether clusters are ``standard candles''.

\subsection{Testing for the Origin of Scatter}
\label{sec:scatter2}
A striking feature of the FP in our fiducial model, shown in
Figure~\ref{fig:hyperplane}, is its orientation, which implies that
the three projections have significantly different scatter.  In
particular, a relatively large scatter is predicted around the
($\theta-\Delta S_\nu$) and ($\theta - T_X$) scaling relations, while
the ($\Delta S_\nu - T_X$) relation remains much more tightly defined.
Figures~\ref{fig:SZthall} and \ref{fig:SZ-theta} reveal that this
result follows from the deterministic nature of our fiducial model,
where the only source of scatter in the observables is the
distribution in formation redshift $z_f$ of clusters (which,
ultimately, is a direct consequence of cosmological initial
conditions).  As discussed in \S~\ref{sec:scatter} above, the
deterministic scenario, without any additional sources of scatter,
might be consistent with existing data (i.e. the scatter in the
$\theta - T_X$ and $M-T_X$ relations).  Nevertheless, it is
interesting to consider alternative options, those in--between the two
extreme {\it deterministic} and {\it stochastic} scenarios.

The parameters $x$, and $f_M$, introduced in \S~\ref{sec:scatter}
above, quantify how much of the observed scatter is caused by
cosmological initial conditions, or by random variations from cluster
to cluster, arising from their internal physics. To quantify this we
Monte--Carlo generate the $M-T$ relation of a mock catalog of about
250 clusters, first in our deterministic fiducial model with
($f_M=0.75$, $x=0$), and then for different combinations of $f_M$ and
$x$. We then compare those to the fiducial model with a 2 dimensional
Kolmogorov-Smirnov test (e.g., \cite{NUMREC,Peacock83,FF87}). This test
computes the $D$ statistic, that is the maximum cumulative difference
between the two distributions, and then gives $P_{>D}$, the
probability that $D$ could be greater than observed. 
Small values of
$P_{>D}$ indicate that it is extremely likely that the two
distributions differ, for example a conservative choice is usually
$P_{>D}<0.01$ which implies that the two data sets are
significantly different. We have checked with multiple realizations that if
two datasets are drawn from the same underlying distribution (i.e. our
fiducial model) $P_{>D}>0.317$  about 68\% of the times and $P_{>D}>0.046$
about 95\% of the times. In what follows we will thus plot the equal
probability contours corresponding to $P_{>D}=0.317, 0.046$ and  $0.01$.

Figure~\ref{fig:KSMT} show these equal probability contours. 
Note
that the {\it deterministic} scenario corresponds to ($f_M=0.75;x=0$);
the {\it stochastic} scenario corresponds to ($f_M=1;x>0$); and
in-between cases are described by ($f_M<1;x>0$).

\myputfigure{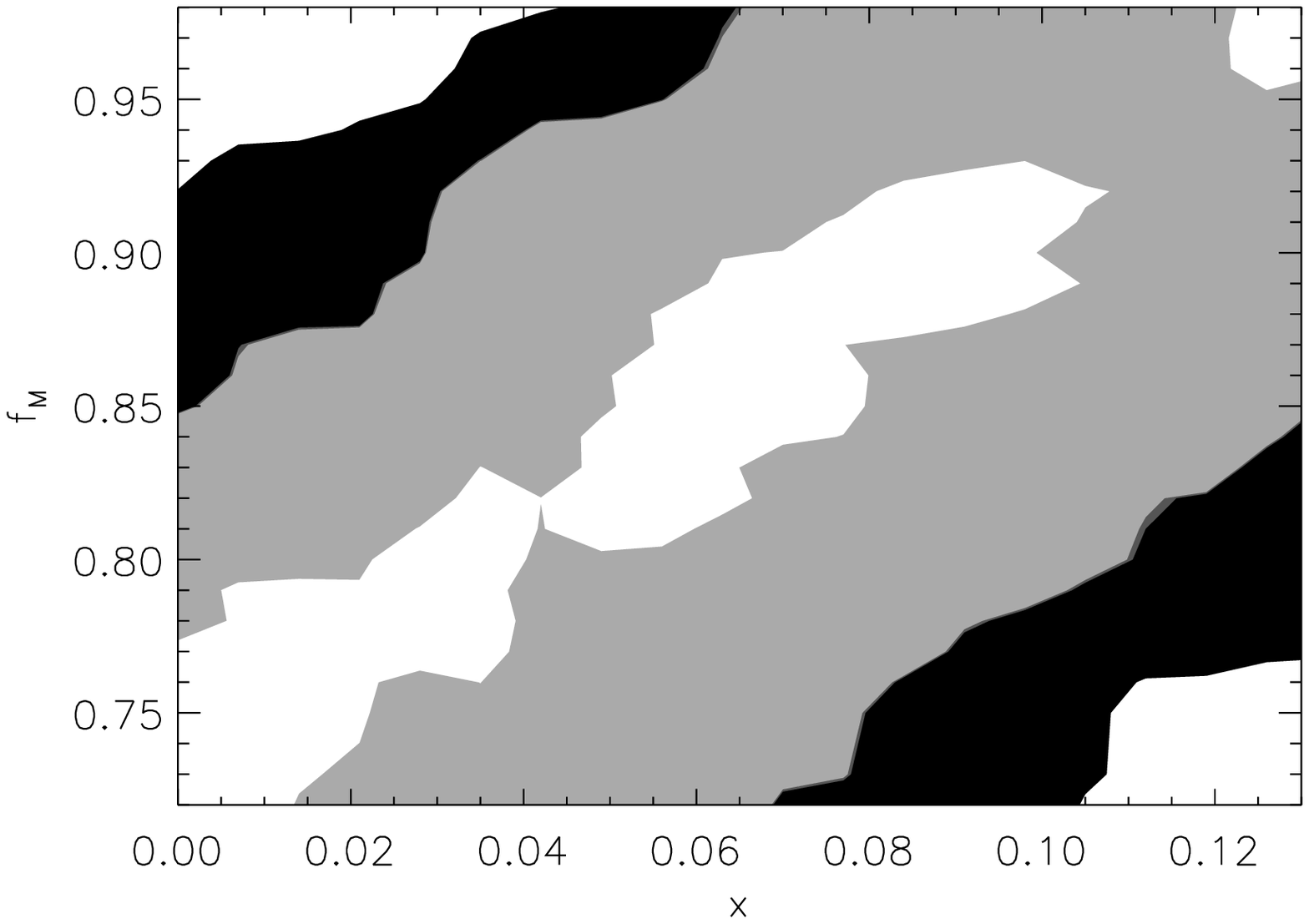}{3.2}{0.50}{-15}{0}
\figcaption{\label{fig:KSMT}
We illustrate a probabilistic study of the origin of the scatter in
the $M-T_X$ relation.  The scatter can arise from cosmological initial
conditions ($f_M=0.75,x=0$), or from a stochastic source
($f_M=1$). Other choices of ($f_M,x$)--values correspond to a
combination of both effects.  The equal--probability contours for
$P_{>D}=0.317, 0.046, 0.01$, where $P_{>D}$ is the likelihood that the
two data sets are drawn from the same distribution, are shown in the
$f_M$-$x$ plane. The likelihoods were obtained from a 2D
Kolmogorov-Smirnov (KS) test of the $M-T_X$ distribution of 250
clusters at redshift $z=0$.  Probabilities are obtained by comparing
the $M-T_X$ distributions to that in our fiducial flat $\Lambda$CDM
model with $f_M=0.75$ and $x=0$.  The parameter space outside the light  gray
area is excluded by the KS test at $\sim$ 68\% confidence. }
\vspace{\baselineskip} 

Figure~\ref{fig:KSMT} has two important implications. First, the
parameter space outside the light gray area is excluded by the KS
test at the 68\% confidence level ($P_{>D}<0.317$) , indicating that the $M-T_X$ relation by itself can already be a
useful discriminant for the source of scatter.  Forthcoming datasets
can yield a local sample of many more than 250 clusters with size,
temperature, and mass estimates, in which case the constraints shown
in Figure~\ref{fig:KSMT} can be significantly improved.  Second,
Figure~\ref{fig:KSMT} suggest that although their combination is
tightly constrained from the $M-T_X$ distribution alone, a degeneracy
appears to still remain between $f_M$ and $x$.  A comparison with
Figure~\ref{fig:hyperplane} suggest that this degeneracy can be broken
by considering, in addition, the distribution of clusters along the
$\Delta S_\nu - T_X$ relation.  For models where the scatter is
cosmological ($f_M\approx 0.75$), this relation is significantly
tighter than the $\theta-T_X$ distribution.  On the other hand, in
``stochastic'' models ($f_M\approx 1$), the scatter in $\Delta S_\nu -
T_X$ should be comparable to that in $M - T_X$.

\myputfigure{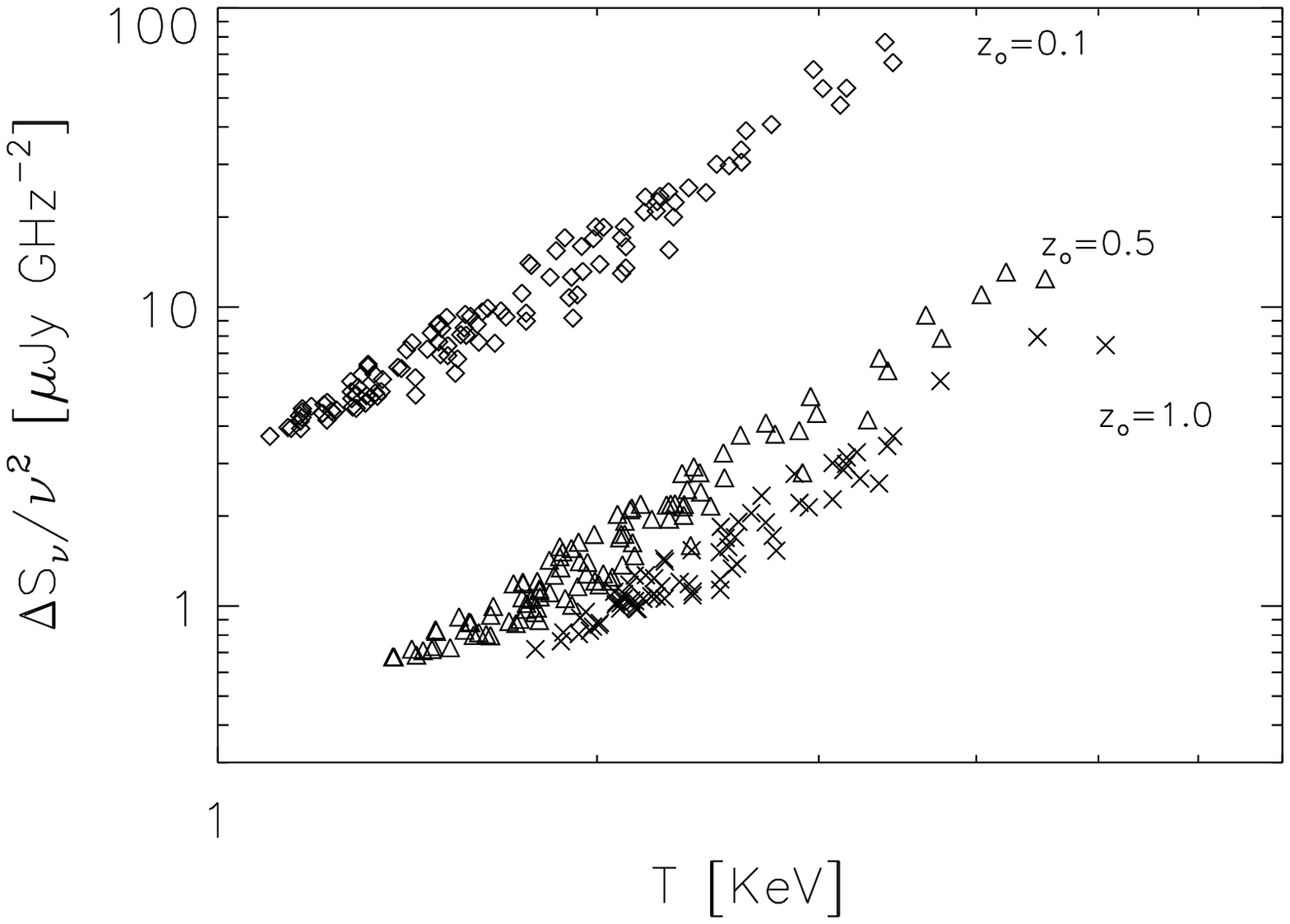}{1.2}{0.25}{-25}{+10}
\myputfigure{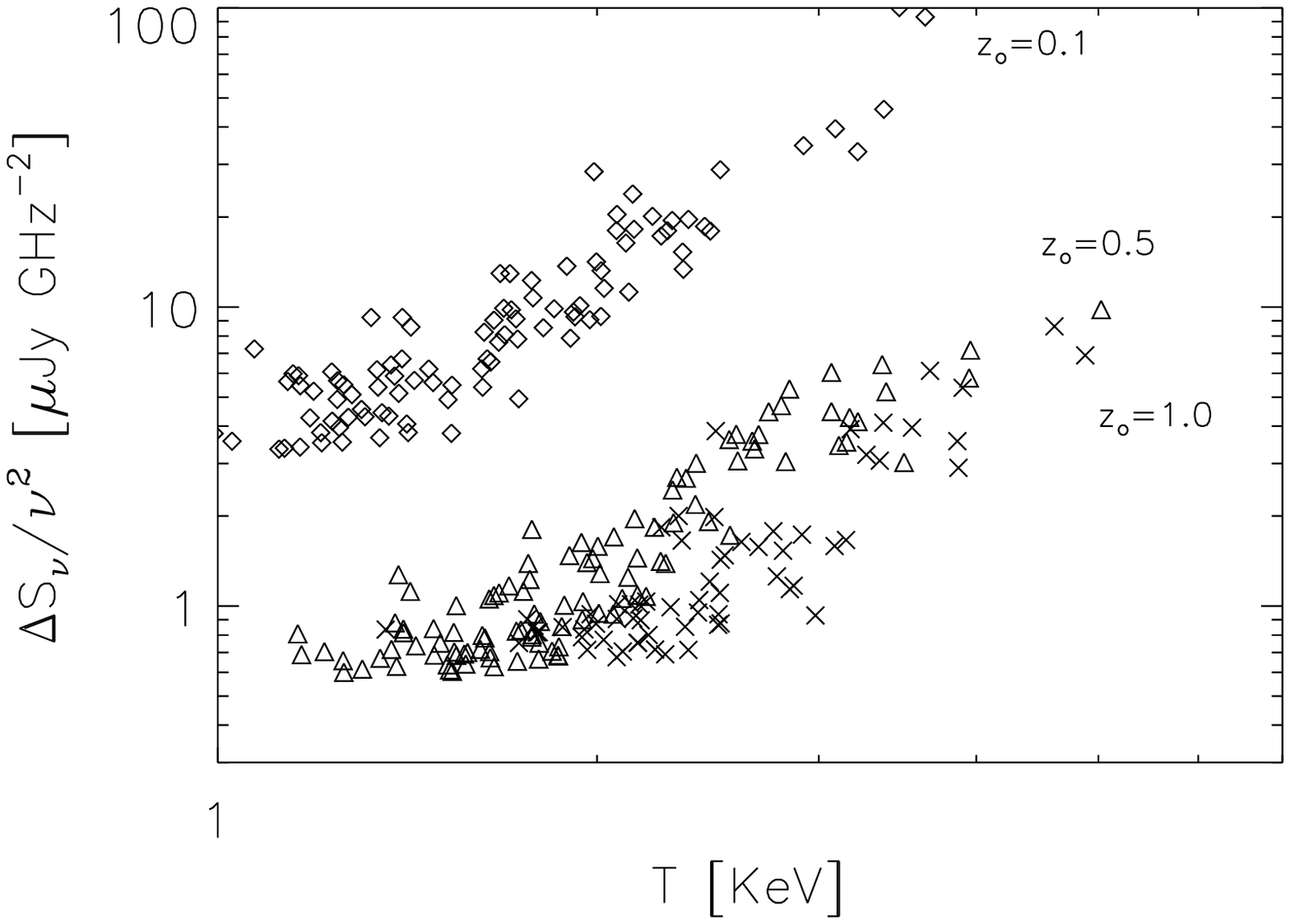}{5.0}{0.25}{-10}{-83}
\figcaption{\label{fig:SZzi3}
The distribution of 100 Monte Carlo clusters in the $\Delta S_\nu-T_X$
plane. Clusters were selected with masses above $10^{14}h^{-1}$
M$_\odot$, observed at redshifts $z_o=0.1$ (diamonds), $z_o=0.5$
(triangles), and $z_o=1$ (crosses). The left panel corresponds to our
fiducial model, in the {\it deterministic} scenario, where the scatter
is purely from cosmological initial conditions. The right panel shows
the {\it stochastic} scenario , with a stochastic scatter. Although
both scenarios, by construction, have the same scatter in the $M-T_X$
relation, the stochastic scenario predicts a significantly larger
scatter in $\Delta S_\nu-T_X$ than the deterministic one.}
\vspace{\baselineskip} 

We therefore next examine constraints on ($f_M,x$) that will be
available from clusters with measured $z$ and $\Delta S_\nu$.  In
order to illustrate how the scatter in the $\Delta S_\nu-T_X$ relation
differs in the stochastic and deterministic scenarios, in
Figure~\ref{fig:SZzi3} we show the distribution of clusters in this
plane.  We used Monte-Carlo realizations of our fiducial model to
generate a catalog of $\sim 100$ clusters with masses above
$10^{14}h^{-1}$ M$_{\odot}$ (cf.~\cite{Holderetal00}), observed at
each of the redshifts $z_o=0.1$, $0.5$, and $1$.
 
The left and right panels of Figure~\ref{fig:SZzi3} show the
distributions in the deterministic and stochastic scenarios,
respectively.  A comparison of the two panels reveals that at any
redshift, the scatter in the deterministic scenario is small; while in
the stochastic scenario, it is significantly larger.  We find that the
two scenarios differ increasingly towards higher redshifts. This is
not surprising, since in the deterministic scenario, the scatter is
reduced by the narrower distribution of cluster ages at high--$z$,
while in the stochastic scenario, there is no similar trend.

In what follows, we find it useful to define a combination $\eta$ of
the observables ($\Delta S_\nu,T,z$) by
\be
\eta\equiv \frac{2.45\times10^{-4}(T_X/[KeV])^{5.532}}
{(1+z)\Omega_0 \wp(z)(\frac{\Delta S_{\nu}}{\nu^2}/[\mu {\rm Jy}/{\rm GHz}^2] f_M d_A^{\prime 2})^2}
\label{eq:h}
\ee
Under the assumption that a cluster at redshift $z$ formed at the same
redshift $z_f=z$, $\eta$ would simply equal the Hubble constant
$\eta\equiv h$.  In general, for a sample of clusters with different
ages, $\eta \geq h$ since $z_f\geq z$.  As a result, in our fiducial
model ({\it deterministic} scenario), for any given redshift or
temperature, the lowest value of $\eta$ ($\eta_{min}$) in a cluster
sample gives an estimate of $h$.  In the left panel of
Figure~\ref{fig:multiHzo}),
we show the distribution of clusters in the $\eta-z$ plane.
Inadequacies of our fiducial model will show up as a systematic
dependence of $h$ ($\eta_{min}$) on $z$ (and/or on $T_X$; see further
discussion below).

We next quantify how the combination $\eta$, defined in
equation~(\ref{eq:h}), depends on $z$ and $T_X$ in different models.
We generate a Monte-Carlo mock catalog of $\sim 300$ clusters between
redshift $0\leq z\leq 1$ in our fiducial model.  We impose the
restriction $z\leq 1$ because of the increasing difficulty to obtain
accurate cluster redshifts for more distant clusters.  The survey
proposed by Holder et al. (2000) will yield about $300$ clusters with
$z\lap 1$.  Similarly, in the CSCS survey proposed by Page et
al. (2001), about $\sim 300$ of the detected SZE clusters will have
redshifts measured in X-rays and/or optical follow--up observations.
For each cluster in our mock catalog, we then compute $\eta$, using
equation (\ref{eq:h}).

\myputfigure{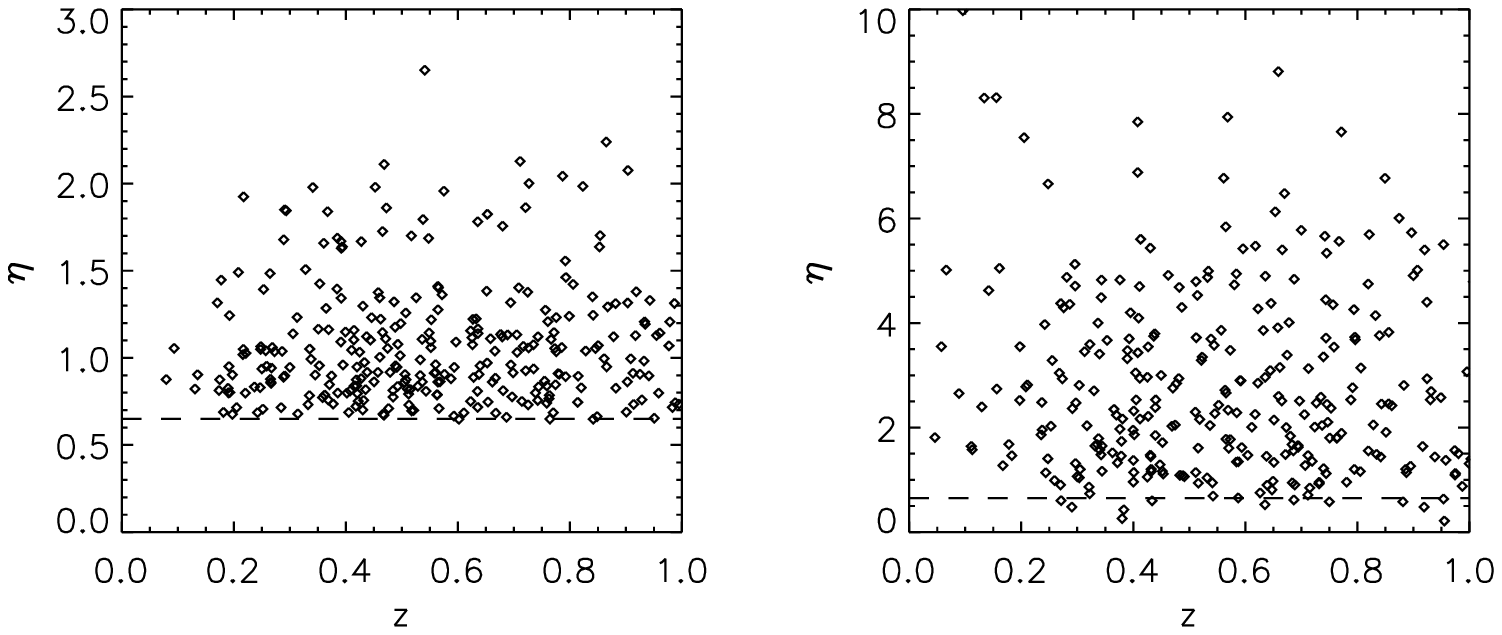}{3.2}{0.50}{-25}{-00}
\figcaption{\label{fig:multiHzo} Scatter plot of $\eta$ from Monte Carlo
simulations of $\sim 300$ clusters. The true underlying cosmological
parameters are those for our fiducial model ($\Omega_0=0.4$,
$\Lambda_0=0.6$, $h=0.65$). The horizontal dashed line show where
clusters for which $z_o=z_f$ should lie. The left panel corresponds
to our {\it deterministic} fiducial model, while the right is
for the {\it stochastic} model. Note the different scale in the y axis
and the different distributions of points relative to the dashed line.}
\vspace{\baselineskip} 

Using these Monte-Carlo catalogs, we can now determine constraints on
($f_M,x$).  We compare mock catalogs of $\sim 300$ clusters, assuming
different combinations of ($x,f_M$) in otherwise identical models.
Instead of using the deterministic scenario for the benchmark model,
we adopt $f_M=0.85$ and $x=0.065$. This combination lies approximately
in the center of the degenerate parameter range shown in
Figure~\ref{fig:KSMT}, and hence it allows us to quantify the
``distance'' of both deterministic and stochastic scenarios from this
intermediate case.  The $P_{>D}$=0.317, 0.046, and 0.01 probability
contours obtained from these KS tests are shown in
Figure~\ref{fig:KSxf-zh}.

\myputfigure{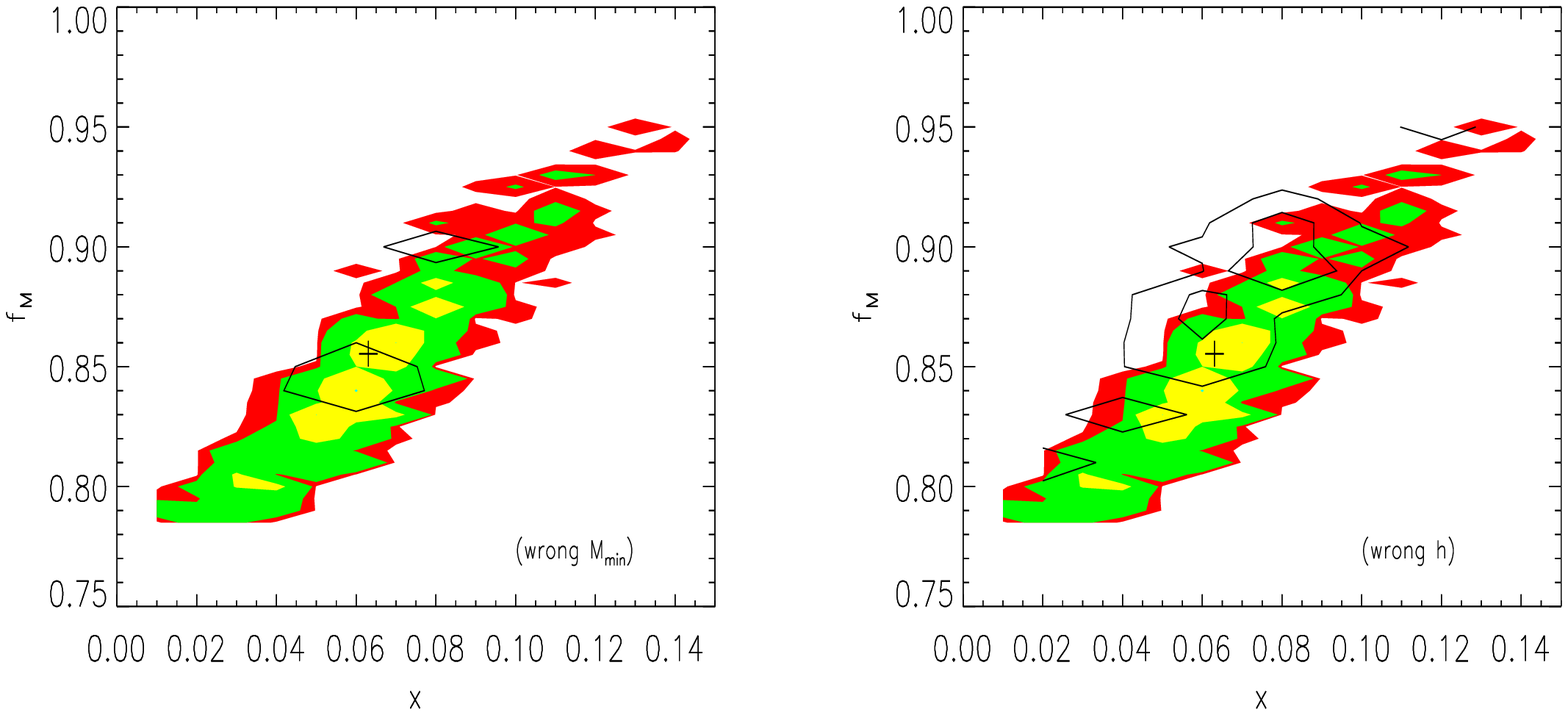}{3.2}{0.50}{-15}{-00}
\figcaption{\label{fig:KSxf-zh}
Equal probability contours at $P_{>D}$=0.317, 0.046 and 0.01 in the
($x,f_M$) plane, obtained from 2D KS tests between distributions of
the clusters in the ($\eta,z$) plane. Each pair of ($x,f_M$) was used
to generate a mock catalog of 300 clusters, and its ($\eta,z$)
distribution was compared to that in a benchmark model with $f_M=0.85$
and $x=0.065$.  The parameter space outside the light gray area is
excluded by the KS test at 68\% confidence.  The two sets of
transparent contours show $P_{>D}=0.317$ and $P_{>D}=0.046$ (nominally 68.3\% and
95.4\% respectively)  contours
obtained by generating mock catalogs down to a factor 2 lower value of
$M_{\rm min}$ than in the fiducial model (left panel); or similarly,
assuming the wrong value of $h$ by +5$\%$ (right panel).  Larger
errors on $M_{\rm min}$ or $h$ would imply that no combination of
($x,f_M$) produces a ($z,\eta$) distribution consistent with our
fiducial model. }
\vspace{\baselineskip} 

The shaded contours (identical on both panels) reveal that tight
constraints can be derived on a combination of ($x,f_M$), similar, but
not identical to those from the $M-T$ relation.  In using these types
of constraints, it is important to compare observational data with
model catalogs extending down to the ``correct'' minimum mass,
i.e. the true mass of the smallest observed cluster.  The shaded
confidence regions were obtained by implicitly assuming the right
minimum mass ($M_{min}$) of the survey.  To demonstrate the
sensitivity of this method to the limiting mass, in the left panel of
Figure~\ref{fig:KSxf-zh}, the set of transparent contours shows
$P_{>D}=0.317, 0.046, 0.01$, obtained assuming a value for $M_{\rm min}$
that is too low by a factor of two.  Similarly, the transparent contours
in the right panel show the results we obtain if we assume the wrong
value of $h$ by +5$\%$.  A larger error on these two parameters would
results in all combinations of ($x,f_M$) producing a ($z,\eta$)
distribution that is inconsistent with the distribution in our
benchmark model. This conclusion does not change if we reverse the sign of the
change in $M_{min}$ and $h$. This suggests that this type of study can be used to
simultaneously constrain ($x,f_M$), as well as $M_{\rm min}$ and $h.$
Our results in Figure~\ref{fig:KSxf-zh} also suggest that if one
marginalizes over the allowed ranges of $M_{\rm min}$ and $h$, the
errors on $x$ and $f_M$ will not be significantly increased.

\subsection{Cluster Evolution and Feedback}
\label{sec:notophat}

In \S~\ref{sec:clusterphysics} we introduced two parameters that
describe deviations arising from energy injection, or other feedback
($\xi$); and from lack of full--virialization, or redshift--evolution
due to other reasons ($\alpha$).  In our fiducial model, $\xi=1.5$ and
$\alpha=1$.  It is interesting to quantify the effects on the cluster
scaling relations of different choices for both parameters.  As
suggested in the previous section, an inadequacy in our fiducial model
will show up as a systematic dependence of $\eta$ (eq. 13) on $z$
and/or $T_X$. For example, a 20\% variation in the value of $\xi$ has
a sizeable effect in the distribution of $\eta$ vs $T$, as we
demonstrate in the right panel of Figure~\ref{fig:multiHzo2}.

\myputfigure{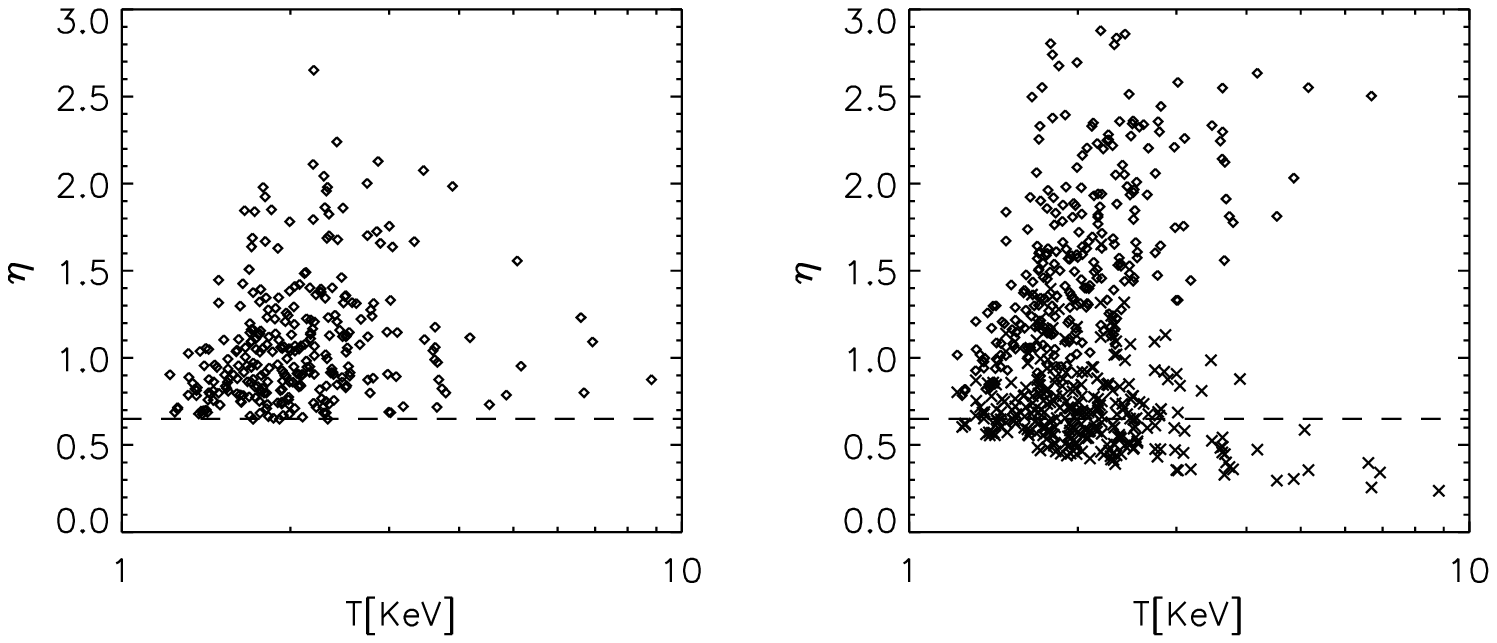}{3.2}{0.50}{-25}{-00}
\figcaption{\label{fig:multiHzo2} Scatter plot of $\eta$ from Monte Carlo
simulations of $\sim 300$ clusters. The true underlying cosmological
parameters are those for our fiducial model ($\Omega_0=0.4$,
$\Lambda_0=0.6$, $h=0.65$). The horizontal dashed line show where
clusters for which $z_o=z_f$ should lie. The left panel corresponds
to our {\it deterministic} fiducial model, while in the right panel  the parameter $\xi$ in the scaling relation
(eq.~\ref{eq:TzfM}) has been changed from the fiducial value $\xi=1.5$
to $\xi=1.8$ (diamonds) and $\xi=1.2$ (crosses).}
\vspace{\baselineskip} 

In this figure, we show $\eta$ vs. $T$ for $\xi=1.5$ (left panel), and
for $\xi=1.8$ (right panel; diamonds) and $\xi=1.2$ (right panel;
crosses), in the fiducial model.

In order to quantify the minimal level of deviations that can be
measured in future SZE surveys, we performed a 2D KS test in the
$\eta-z$ distribution for the $\alpha$ parameter using Monte Carlo
simulations of 300 clusters.  We find that deviations from $\alpha=1$
by $\pm 0.03$ are detectable at 68\% confidence level.  Performing the
same test in the $\eta-T_X$ plane for the $\xi$ parameter, we find
deviations from the value $\xi=1.5$ by $\xi=1.5^{+0.02}_{-0.05}$ are
detectable at the same significance.

If angular sizes of $\sim 300$ clusters in a survey are available, in
addition to ($\Delta S_\nu,T,z$), the values of $x$, $f_M$ and
cosmological parameters can be accurately determined.  From
equation~(12) we obtain:

\be
\frac{h}{d_A'}=\frac{\frac{\Delta S_{\nu}}{[{\rm \mu Jy}]} f_M}{(\frac{\nu}{[{\rm GHz}]})^2f_{\rm ICM}(\frac{T_X}{[{\rm KeV}]})^2 \frac{\theta}{[{\rm deg}]} (1+z_o) 15.6}
\label{eq:allobs}
\ee
If we consider this combination of observables as a function of $z_o$,
we find that (a) the scatter around this relation is not sensitive to
$f_M$, and hence it can only be due to a non-zero value for the
parameter $x$; and (b) the absolute normalization of the relation
depends on the angular diameter distance and $f_M$. In particular, if
cosmological parameters are known, the latter depends only on $f_M$.

\myputfigure{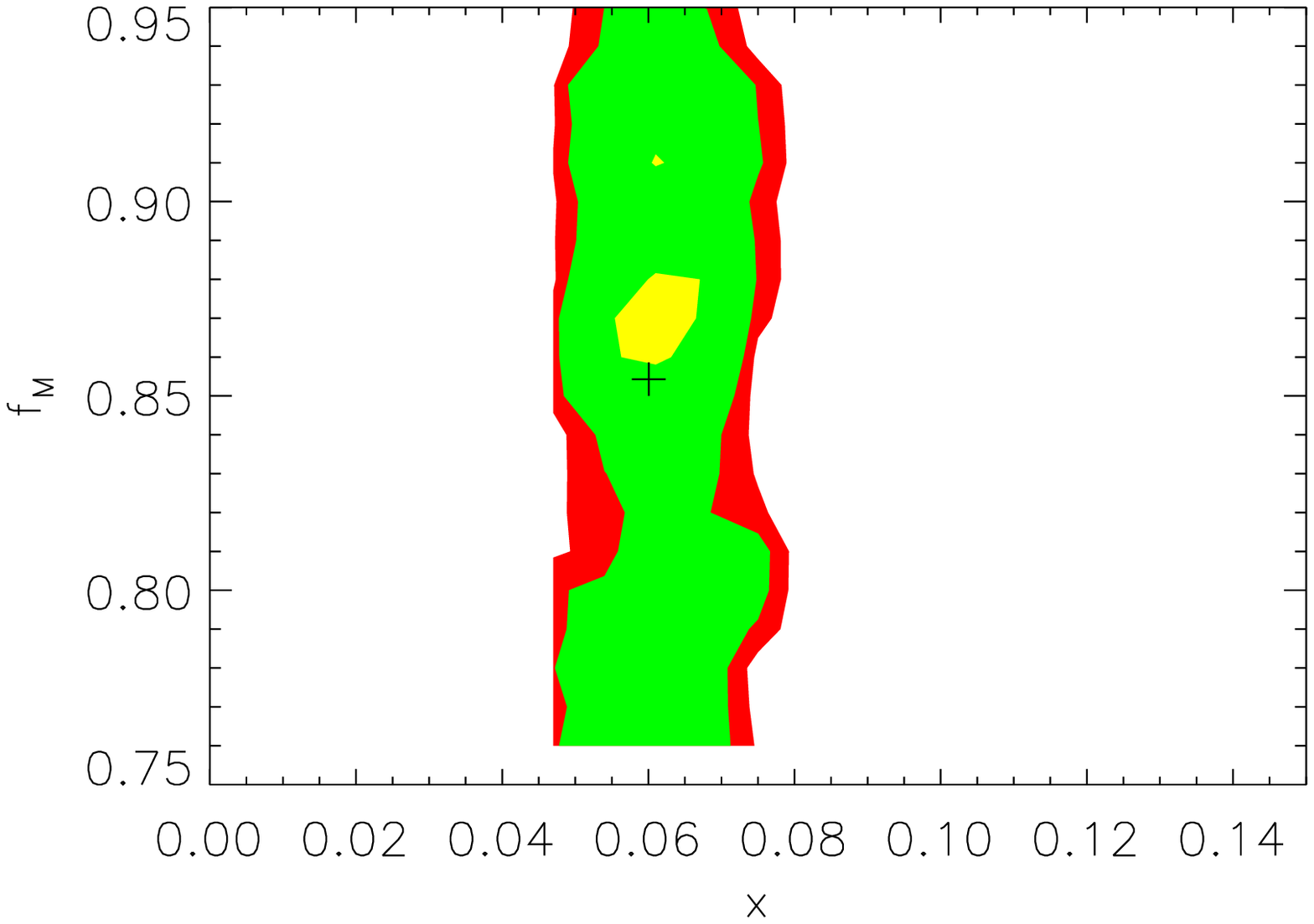}{3.2}{0.50}{-15}{-00}
\figcaption{\label{fig:ks-xf-all}
Equal probability contours at $P_{>D}$=0.317, 0.046 and 0.01 in the
($x,f_M$) plane, obtained from 2D KS tests between distributions of
the clusters using the combination of four observables as in
equation~(\ref{eq:allobs}) as a function of $z_o$.  The combination of
these contours with those of Figure~\ref{fig:KSxf-zh}, allows a unique
combination of ($x,f_M$) to be determined.}
\vspace{\baselineskip} 

These properties of the combination of observables in
equation~(\ref{eq:allobs}) allows a unique combination of ($x,f_M$) to
be measured. Figure~\ref{fig:ks-xf-all} shows the equal probability
contours ($P_{>D}$=0.317, 0.046, 0.01) in the $f_M-x$ plane for a sample
of 300 clusters: the
``product'' of the probability contours of Figures~\ref{fig:KSxf-zh}
and \ref{fig:ks-xf-all} allows a unique combination of ($x,f_M$) to be
determined.  In addition, the redshift dependence of
equation~(\ref{eq:allobs}) directly probes the redshift evolution of
the angular-diameter distance, and therefore can be further utilized
to measure cosmological parameters, as we will discuss in \S~6.3.

\section{Probing Cosmological Parameters}
\label{sec:cosmology}

In this section, we will assume that clusters can be used as standard
candles, i.e. that our modeling of cluster physics is accurate. The
deformations of the FP relative to its shape in the fiducial model can
then be used to determine cosmological parameters.

\myputfigure{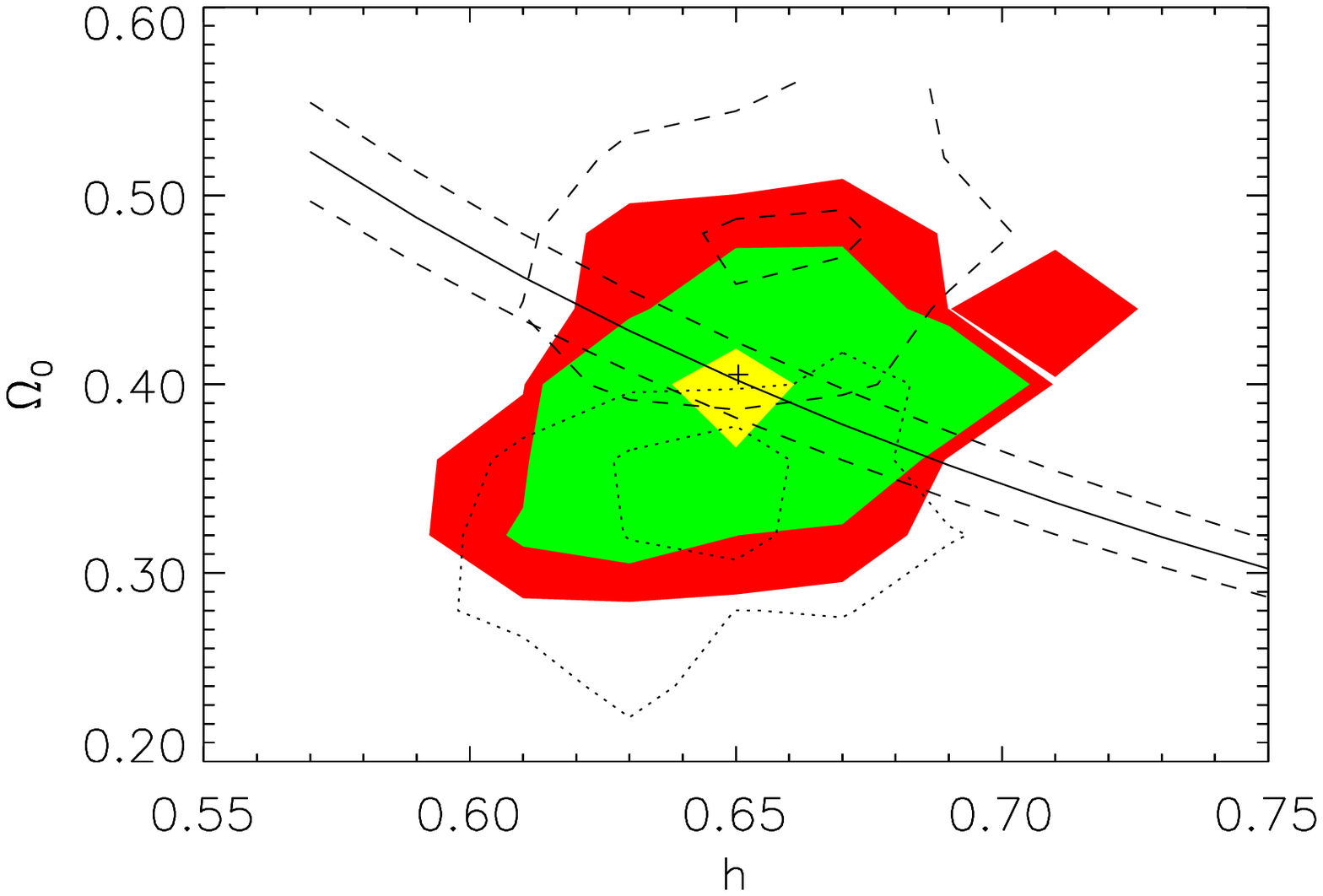}{3.2}{0.50}{-25}{-00}
\figcaption{\label{fig:KShom-zh}
Equal probability contours in the ($h-\Omega_0$) plane obtained from
the 2D KS test of the $\eta-z$ distribution, similar to those shown in
Figure~\ref{fig:multiHzo}. The parameter space outside the light gray
area is excluded by the KS test at the $\sim$ 68\% level
($P_{>D}=0.317$), the other two contours correspond to $\sim 95$\% and $\sim
99$\% confidence levels.  The ``true'' underlying value is indicated by
the cross. The dotted and dashed contours show the first two
likelihood contours ($P_{>D}=0.317$ and $0.046$), obtained by assuming a
minimum mass for the survey $M_{\rm min}$ two times lower and two
times higher than the $10^{14}h^{-1}$ M$_\odot$ used in the mock
catalog. The estimates of $h$ are not strongly affected by the choice
of $M_{\rm min}$. In practice, one will marginalize over $M_{\rm
min}$: this will increase the uncertainty in the recovered $\Omega_0$,
but not in $h$. The solid line, together with the two dashed lines,
show the constraint $\Omega_0h^2=$constant, and $\pm 5\%$
uncertainties, expected from forthcoming CMB experiments.}
\vspace{\baselineskip}

\subsection{Constraints from $\Delta S_\nu,T$, and $z$.}
\label{sec:cosmonotheta}

Assuming only $\Delta S_\nu$ and $T_X$ and $z_0$ are available, a 2D
KS test applied to the $\eta-z$ distribution (see
Figure~\ref{fig:multiHzo}), can be used to constrain $\Omega,h$.  It
is important to note here that this measurement of the Hubble
constant, based on a combination SZE and X-ray data, is different from
an existing method proposed by e.g., Gunn (1978), Silk \& White
(1978), Birkinshaw (1979). The latter method, by combining the central
SZ decrement with X-ray central temperature, yields an estimate of the
length of the cluster; assuming that the cluster is spherical this can
be used as an angular diameter distance test. By averaging over a
large cluster sample, effects of cluster asphericity can be averaged
out. In this method clusters are used as {\it standard rulers}. The
method presented here is complementary; it uses the total observed SZ
flux decrement, which can be directly measured, requires data that is
easier to obtain (no detailed SZ and X-ray map of the cluster are
needed) and by making use of the whole cluster SZD, should therefore
be less sensitive to the details of cluster physics near the
center. This method relies on different assumptions; it assumes
clusters are virialized (although deviations from virialization can be
parameterized, as discussed in \S~2.3 and 2.4). In our method,
clusters are used as {\it standard candles}.  The expected statistical
errors for a cluster sample of the same size are comparable for the
two methods, but the possible systematics are of entirely different
nature.

Results of the 2D KS test of the $\eta-z$ distribution of $\Omega,h$
are shown in Figure~\ref{fig:KShom-zh}.  The parameter space outside
the light gray area is excluded by the KS test at the 68\% level
($P_{>D}=0.317$), the other two contours correspond to $P_{>D}=0.046$ and
$P_{>D}=0.01$.
 The ``true'' underlying value is indicated by
the cross. The dotted and dashed contours show the first two
likelihood contours ($P_{>D}=0.317$ and $0.046$), obtained by assuming a
minimum mass for the survey $M_{\rm min}$ two times lower and two
times higher than the $10^{14}h^{-1}$ M$_\odot$ used in the mock
catalog. The estimates of $h$ are not strongly affected by the choice
of $M_{\rm min}$. In practice, one will marginalize over $M_{\rm
min}$: this will increase the uncertainty in the recovered $\Omega_0$,
but not in $h$. The solid line, together with the two dashed lines,
show the constraint $\Omega_0h^2=$constant, and $\pm 5\%$
uncertainties, expected from forthcoming CMB experiments.  This
suggests that the two constraints are sufficiently different, and can
be combined together to break the degeneracy between $\Omega_0$ and
$h$.

\subsection{Constraints from $\Delta S_\nu,\theta$, and $z$.}
\label{sec:cosmonoT}

\myputfigure{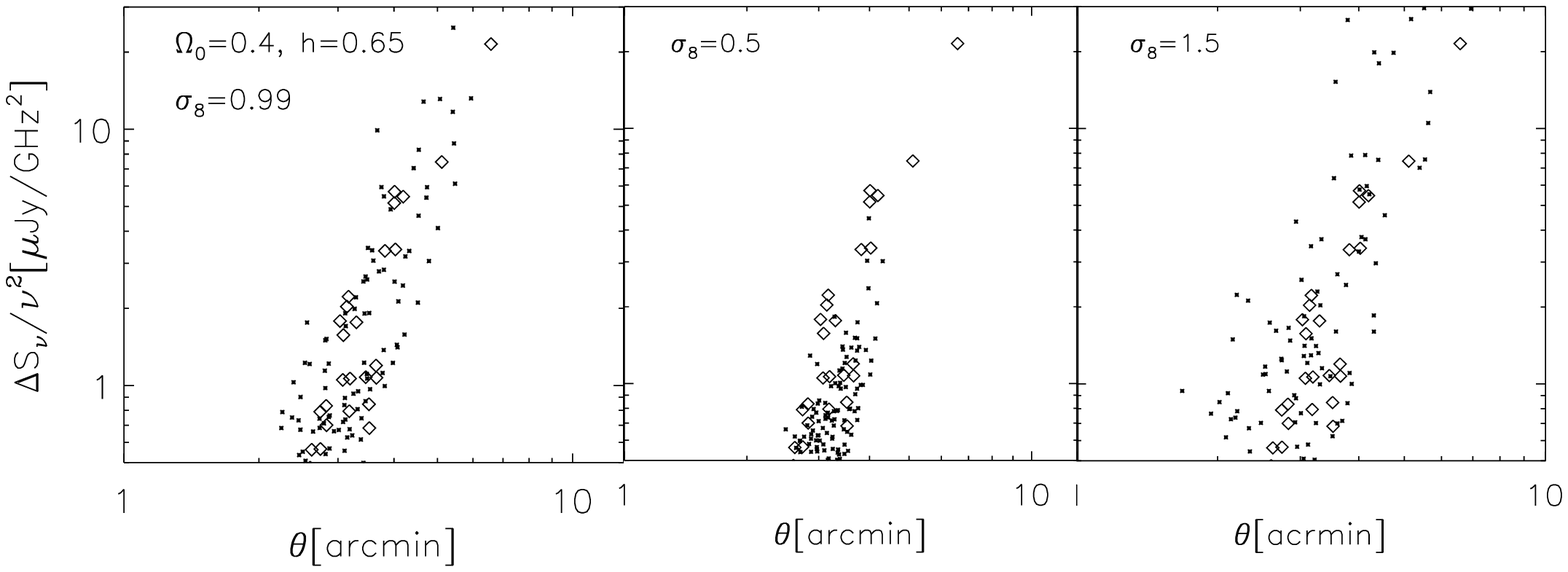}{3.2}{0.50}{-25}{+10}
\figcaption{\label{fig:SZthetaVKMB}
The dependence of the $\Delta S_\nu-\theta$ distribution on $\sigma_8$
($\Delta S_\nu/\nu^2$ is in units of $\mu$Jy/GHz$^2$). The diamonds
show a Monte Carlo sample of 25 clusters at $z=0.3$ with $\theta \gap
2'$ (close to the specifications of the CSCS survey) for
$\Omega_0=0.3$, $\Lambda_0=0.7$, $h=0.65$ and $\sigma_8=0.99$; the
dots show the $\Delta S_\nu$-$\theta$ distribution of $\sim 100$
Monte-Carlo simulated clusters for different choices of $\sigma_8$.}

\myputfigure{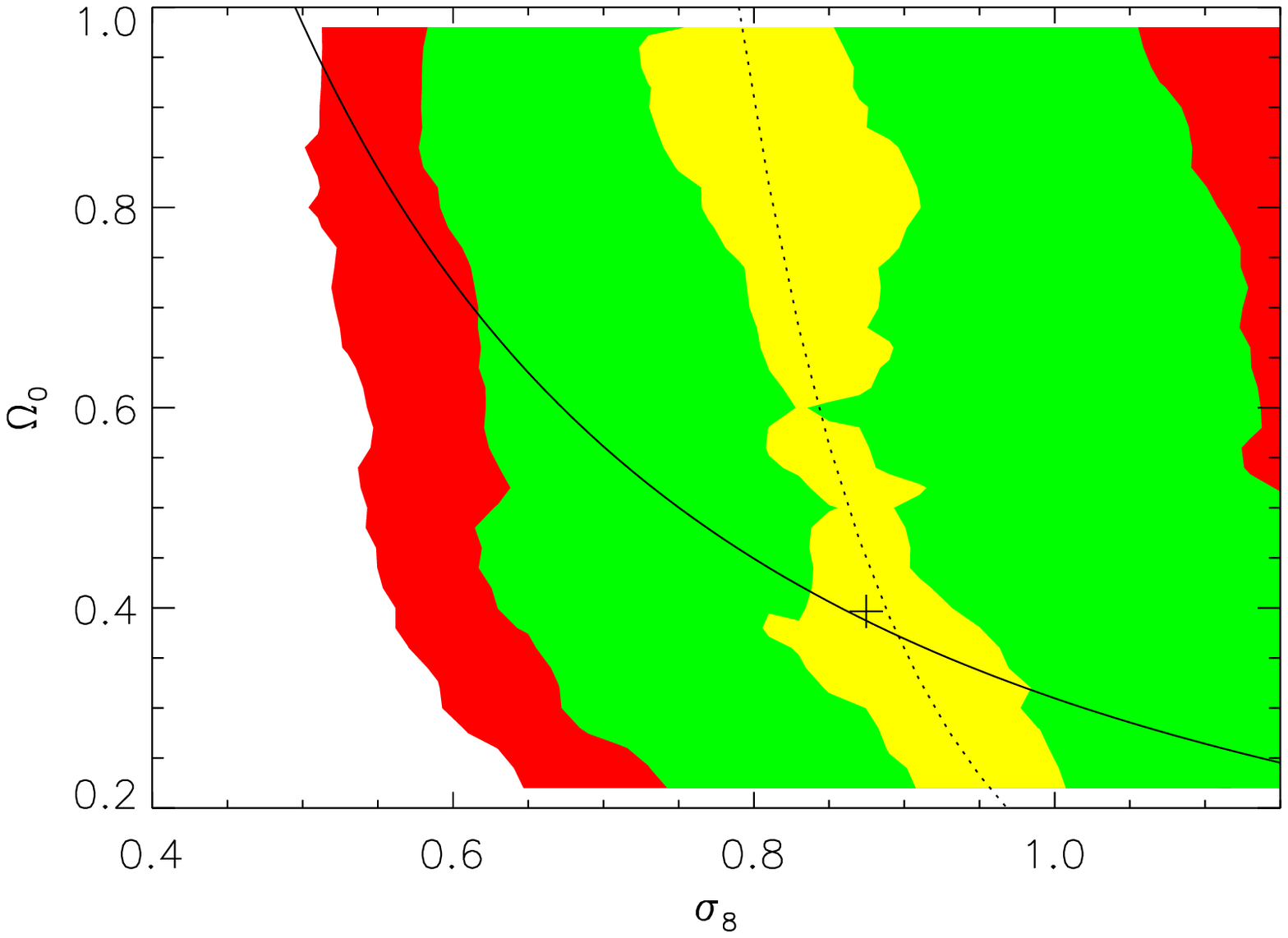}{3.2}{0.50}{-25}{-00}
\figcaption{\label{fig:kssigmom}
Equal probability contours in the $\sigma_8-\Omega_0$ plane, obtained
from the $\Delta S_\nu-\theta$ relation. The parameter space outside
the light gray area is excluded at the 68\% ($P_{>D}=0.317$) level by
the KS test. The constraint we obtain can be approximated by
$\sigma_8=0.92\Omega_0^{-0.11}$. For reference, the solid line shows
the constraint in the $\sigma_8-\Omega_0$ plane obtained from the
local abundance of massive clusters. A comparison suggests that the
two constraints can be used together to measure $\Omega_0$ and
$\sigma_8$ simultaneously.}
\vspace{\baselineskip} 

Assuming that clusters in the sample are resolved, and that their
angular sizes ($\theta$) are measured, a 2D KS test applied to the
$\Delta S_\nu-\theta$ distribution (see Figure~\ref{fig:SZthetaVKMB}),
can be used to constrain $\Omega_0$ and $\sigma_8$. In
Figure~\ref{fig:SZthetaVKMB}, we illustrate the dependence of the
$\Delta S_\nu-\theta$ distribution on $\sigma_8$ ($\Delta S_\nu/\nu^2$
is in units of $\mu$Jy/GHz$^2$). The diamonds show a Monte Carlo
sample of 25 clusters at $z_o=0.3$ with $\theta
\gap 2'$ for $\Omega_0=0.4$, $\Lambda_0=0.6$, $h=0.65$ and
$\sigma_8=0.99$; the dots show the $\Delta S_\nu$-$\theta$
distribution of $\sim 100$ Monte-Carlo simulated clusters for
different choices of $\sigma_8$.
It is clear that a too large/small $\sigma_8$ gives rise to a too large/small
scatter in the ($\Delta S_{\nu}-T$) relation.

The results are demonstrated in Figure~\ref{fig:kssigmom}, where we
show the equal probability contours in the $\sigma_8-\Omega_0$ plane
obtained from the $\Delta S_\nu-\theta$ relation. The fiducial model
is indicated by the cross (here we have dropped the constraint
$\sigma_8=0.495\Omega_0^{-0.6}$). As before, the parameter space
outside the light gray area is excluded at the 68\% level by the KS
test, and the other two sets of contours correspond to 95\% and 99\%
confidence levels.  The constraint obtained from the $\Delta
S_\nu-\theta$ relation can be approximated by
$\sigma_8=0.92\Omega_0^{-0.11}$. For comparison, the solid line shows
the constraint in the $\sigma_8-\Omega_0$ plane obtained from the
local abundance of massive clusters. A comparison of these two methods
suggests that the two constraints can be used together to measure both
$\Omega_0$ and $\sigma_8$.

\subsection{Constraints from $\Delta S_\nu,\theta$, $T$ and $z$.}
\label{sec:cosmo}
As already mentioned in \S~5.2 above, if all observables are known for
a cluster sample, and if $f_M$ and $x$ are reasonably well
constrained, then equation~(\ref{eq:allobs}) yields the angular
diameter distance as a function of redshift.  This quantity depends on
$\Omega_0$, $\Lambda$, $h$ and the equation of state parameterized by
$w$; it can thus be used to constrain these parameters. Several other
observables have already been used to measure the angular diameter
distance as a function of redshift, such as type Ia supernovae (e.g., Reiss et
al. 1998; Perlmutter et al. 1998)
or the redshift evolution of the linear power spectrum shape (e.g., McDonald
\& Miralda-Escude 1999; Roukema \&
Mamon 2000).
Galaxy clusters potentially offer an independent test.  The accuracy
with which the angular diameter distance can be measured will strongly
depend on the observational uncertainties in $\Delta S_\nu,\theta$ and
$T$ and will be addressed in a subsequent work.
 
\section{A Comparison of Two Different Statistics}

Throughout this paper, we have adopted the 2D KS test to quantify the
discriminating power of the scaling relations between different models.  In our
study, we have focused on the {\it shapes} of the scaling relations, and have
not explored constraints from the {\it total number} of detected clusters,
which itself is a function of cosmological parameters.  
The cosmology dependence of the number counts has been recently studied by Haiman, Mohr \& Holder (2001)
and Holder, Haiman \& Mohr (2001) to forecast precision-constraints that will
be available on the equation of state parameter $w$, as well as the parameters
$\Omega_0, \Omega_\Lambda$, and $\sigma_8$.
These constraints are
complementary to those that derive from the shape of the scaling relations, and
the two methods will have different systematic errors.  For example, the
fundamental plane approach requires extensive X-ray/optical follow-up of SZ
surveys, or requires the temperatures of SZ selected clusters to be obtained
from different catalogs.  This can introduce selection effects which affect the
number of detected clusters, but not the shape of the scaling relations.
The KS test has several attractive features, such as its distribution
independence and  the fact that is a robust and more direct method.  It is a particularly
appropriate statistic for the present application, since it distinguishes
distributions based only on their shape, independent of their overall
normalization.  Nevertheless, one must bear in mind that the KS test is
optimized to detect ``shifts'' in distributions, and is less well--suited in
distinguishing among distributions with significant ``tails'' (e.g. Press et
al. 1992). In order to assess the robustness of our results to the choice of
statistic, we here compare our analysis of the constraint on the Hubble
constant $h$ with constraints obtained from a maximum likelihood method.  In
applying both types of statistic, we utilize the distribution of clusters in
the ($\eta - z_o$) plane,
as defined in \S~5.1 above, to derive constraints on the Hubble constant.

In general, a maximum likelihood analysis is not directly comparable to the KS
test, since the former utilizes information from the total number of clusters,
while the latter is insensitive to it.  However, in the case of constraining
the Hubble constant, we find that the total number of clusters is insensitive
to the value of $h$.  As an example, in our fiducial model with $h=0.65$, we
predict, by construction, 300 clusters between redshifts $0.1<z<1$, in a solid
angle of $\sim 20$ deg$^2$, above the limiting mass $M_{\rm min}= 10^{14}h^{-1}
{\rm M_\odot}$.  Assuming that the minimum mass $M_{\rm min}$ scales as $M_{\rm
min}\propto h^{-1}$ (e.g. from mass determinations from X--ray profiles, weak
lensing, etc.), and assuming further that the power spectrum has been
independently measured, we find that the total count is essentially independent
of $h$.  Lack of a-priori knowledge of the power spectrum introduces a small
$h$-dependence through the ``shape parameter'' $\Gamma \sim \Omega h$: we find
297.3 clusters for $h=0.64$.  Finally, assuming a fixed $f_{\rm ICM}$, the mass
corresponding to a constant SZ decrement scales more strongly, as $M_{\rm
min}\propto h^{-8/5}$ (with the $f_{\rm ICM}$ dependence used in this paper,
the dependence would be weaker).  Under this scaling, including the $h$-dependence of the
power spectrum, we find 291.5 clusters for $h=0.64$, implying that based on the
number counts alone, models with $h=0.64$ and $h=0.65$ would only be
0.5$\sigma$ apart.  We conclude that in the case of constraints on $h$, we can
directly compare the likelihood and the 2D-KS test performances, both of which
will test primarily the shapes, rather than the normalizations, of the
underlying distributions.

The observation of a discrete number $N$ clusters is a Poisson process, the
probability of which is given by the product
\be
P=\prod_{i=1}^N(e_i^{n_i}\exp[-e_i]/n_i!)  
\ee  
where $n_i$ is the number of clusters observed in the $i^{\rm th}$ experimental
bin, and $e_i$ is the expected number in that bin in a given model:
$e_i=I(\vec{x}_i)\delta\vec{x}_i$. Here, $I$ is proportional to the probability
distribution, and $\vec{x}$ denotes the set of observables, in the present
case, $\vec{x}=(\eta, z_o)$.  Following Cash (1979), for unbinned data (or equivalently for very small
bins that have only 0 or 1 counts) we define the quantity
\be 
C\equiv -2 \ln P =2(E-\sum_{i=1}^N \ln I_i),
\label{eq:cash}
\ee 
where $N=300$ is the total number of observed clusters (i.e. the number of
clusters in the mock catalog for our fiducial model with $h=0.65$), $E$ is the
expected number of clusters in models with arbitrary $h$.  We have omitted all
terms that involve $\delta \vec{x}_i$ since they do not depend on $h$.  The quantity $\Delta C$, defined
in two models with different $h$, has a $\chi^2$ distribution (e.g. Cash
1979). Both the 2D-KS test and the maximum likelihood test are performed in the
$\eta-z_o$ plane, where $I(\eta,z_o)$ is given by:
\be I(\eta,z_o)\equiv
\frac{d^2N}{d\eta dz_o} 
\ee 
where
\begin{eqnarray}
\frac{d^2N}{d\eta dz_o} &=&
\frac{dn}{d\eta}\frac{dV}{dz_o d\Omega}\Delta\Omega\\ \nonumber
&=&
\left.
\left(\frac{d\eta}{dz_f}\right)^{-1}\!\!
\int_{M_{\rm min}}^{\infty}\frac{dN}{dM}\right|_{z_o}\!\!dM\!\frac{dP}{dz_f}
\frac{dV}{dz_o d\Omega} \Delta\Omega \nonumber.
\label{eq:liketa}
\end{eqnarray}
Here $dN/dM$ is the Press--Schechter mass function, $dP/dz$ is the formation
redshift distribution from Lacey \& Cole (1993), and $dV/dz d\Omega$ is the
comoving cosmological volume. We have assumed a solid angle $\Delta\Omega=20$
deg$^2$, and included clusters at redshifts $0.1<z<1$ in our analysis. Recall
also that
\be
\eta=h\frac{(1+z_f)^3\wp(z_f)}{(1+z_o)^3 \wp(z_o)}\;.
\ee

\myputfigure{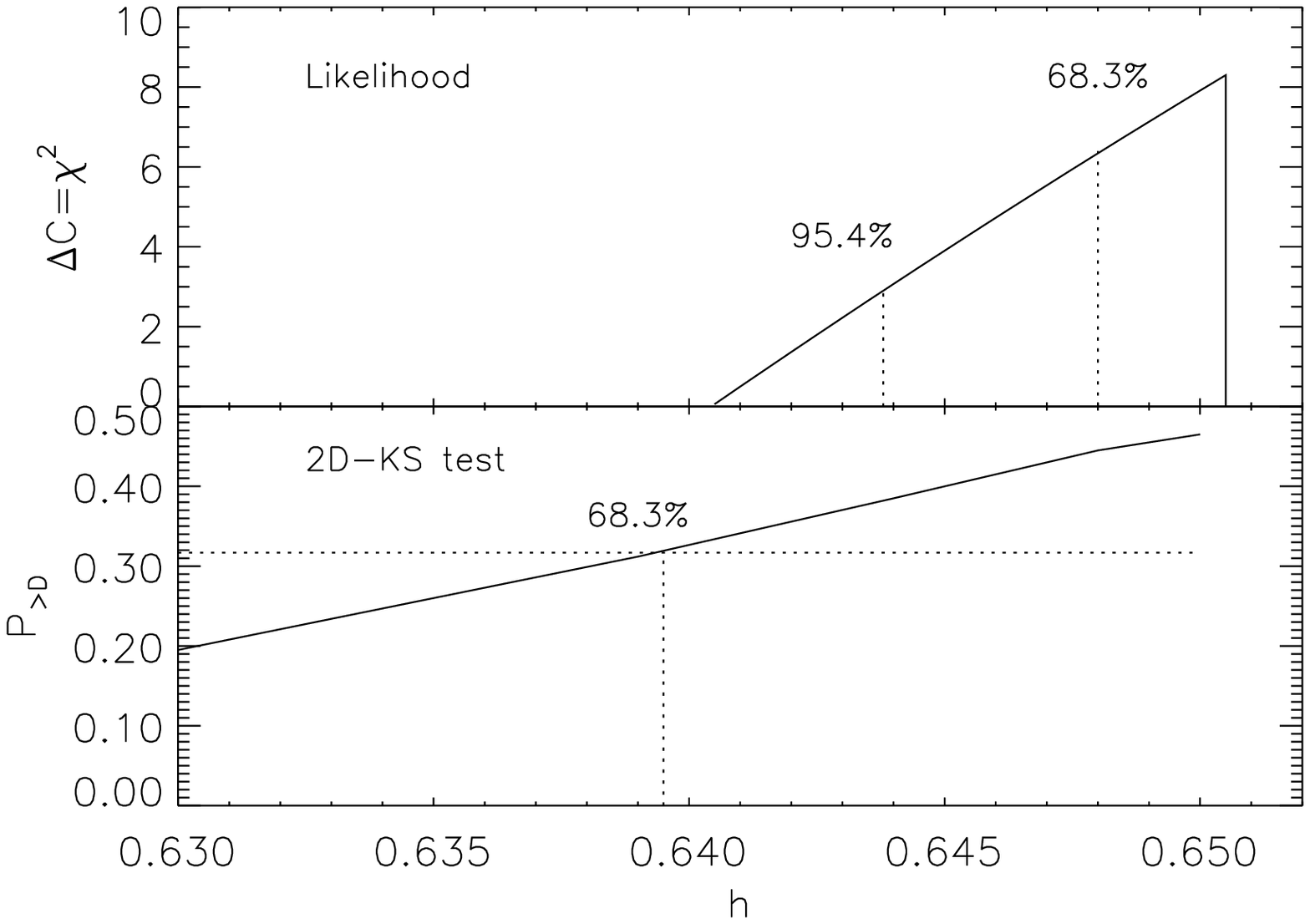}{3.2}{0.50}{-25}{-00}
\figcaption{\label{fig:lik}
Comparison of 2D-KS and maximum likelihood confidence levels. Since the
cluster abundance is insensitive to $h$ it is possible to directly compare the
performances of the two tests for a sample of 300 clusters in out fiducial
(deterministic) model. 
The likelihood is obtained following equations (\ref{eq:cash}) and (\ref{eq:liketa}).
The maximum likelihood method is much more sensitive
that the KS test, but it requires good  knowledge of the cluster mass function
and of mask and selection effects of the survey.}
\vspace{\baselineskip} 

In Figure \ref{fig:lik}, we show the comparison between the likelihoods
obtained from $\Delta C$ and the from the 2D-KS confidence levels, as a
function of $h$.  Since the likelihood is non--Gaussian, the confidence levels
have been obtained by integrating the likelihood curve.  
The figure reveals
that the likelihood method is more sensitive than the KS test.  
For example, in the KS
case, the model with $h=0.64$ is approximately $1\sigma$ away from the fiducial
$h=0.65$ model, while the $\Delta C$ test puts this model at $>2\sigma$
significance. 
This finding suggests that a maximum likelihood method better captures the
shape of the $\eta$ $vs$ $z$ distribution of clusters, and can significantly
improve on the statistical constraints quoted here. We expect that
a similarly improved use of the distribution shapes
may be possible in other cases (e.g., in the measurements
of $\Omega_0$).
However, constraints obtained from a 
likelihood analysis can be dominated by the information
available from the total cluster abundance, rather than the shape.
In those cases, and improved utilization of the shapes can result in
only marginal improvements of the overall constraints, and, at the
same time, the use of the maximum likelihood method would subject 
the results to additional systematic errors.
A systematic study of the best statistic in each case is beyond
the scope of the present paper, and is deferred to a future publication.

\section{Discussion}
\label{sec:discussion}

In anticipation of extensive data on a large sample of galaxy
clusters, it is important to analyze cluster properties to reduce
systematic errors; to test whether physical properties of the local
cluster sample are preserved at high redshift and, if necessary, to
model their evolution.  We present a first attempt to do so
self--consistently from cluster data.

The approach presented here assumes that the primordial fluctuation
field was Gaussian. However, clusters are rare peaks of the density
fluctuation and are therefore extremely sensitive to small deviations
from gaussianity.  Nevertheless, this assumption is not crucial: a
parameterization of non-gaussianity could easily be included in the
model (e.g.,\cite{RGS99}; Matarrese, Verde \& Jimenez 2000; Verde et al. 2000; Verde et al. 2001).
 
We have used four observables: the redshift, the X-ray temperature,
the total observed SZ flux decrement, and the cluster angular size.  A
particular worry in our approach is that we estimate the angular size
based on the virial radius.  Current numerical simulations of cluster
formation (see, e.g., Bryan \& Norman 1998) indeed indicate the
presence of a virialization shock, as expected from the top--hat
collapse model.  However, the shock tends to be weaker, and located at
larger radii, which can make its detection in the SZE maps difficult,
or impossible.  Cluster angular sizes can also be estimated from X-ray
maps, which, however, effectively probe only the core radius.

Alternatively, similar scaling relations could be obtained by
considering other observables, in addition to those we have considered
here.  The ``fundamental plane'' approach can be generalized to
include, for example, the X-ray luminosity, the velocity dispersion,
or estimates of the cluster masses (e.g. from weak lensing). This
opens up many interesting possibilities. The constraints on clusters
physics and cosmological parameters are complementary to constraints
obtained from other observables; they could thus be combined together
not only with e.g., CMB constraints but also with constraints obtained
e.g., from X-ray clusters abundances, SZE cluster abundances, analysis
of kinetic SZE and weak lensing statistics.

\section{Conclusions}
\label{sec:conclusions}

In view of the advent of high-precision cosmology and the expected
avalanche of cluster data available with SZ experiments in conjunction
with X-ray missions, it is vital to be able to extract the maximum
amount of information form these cluster samples.

We have investigated the scientific potential of obtaining total
observed SZ flux decrements ($\Delta S_{\nu}$) in the context of a
well studied cluster sample, for which X-rays and optical follow-up
are available. In particular, for cosmological studies, one would
ideally like to use clusters as ``standard candles": the use of large
future galaxy cluster surveys for cosmological studies is likely to be
limited by the validity of this assumptions, rather than by
statistical uncertainties (Haiman, Mohr \& Holder 2001).

We used a semi--analytic model to study the expected distribution of
galaxy clusters in redshift ($z$), virial temperature ($T$),
Sunyaev--Zeldovich decrement ($\Delta S_\nu$), and angular size
($\theta$).  In the simplest models, clusters are identified with
virialized, spherical halos.  In this case, at every redshift,
clusters define a ``fundamental plane'' (FP) in the three dimensional
parameter space ($T,\Delta S_\nu, \theta$).  The FP and its
redshift--evolution are sensitive to both the internal evolution of
clusters, and to the underlying cosmological parameters, and can be
used to obtain useful constraints on both.  We have parameterized
possible deviations from this model to include effects of energy
injection or feedback, stochastic scatter in the observables
($T,\Delta S_{\nu},\theta$), and deviations from virial
equilibrium. We have shown that their effect is to create measurable
deformations in the FP.

We have thus derived predictions for clusters scaling relations that
involve the SZ decrement, and studied how these scaling relations
depend on assumptions about the cluster physics and structure (i.e. on
the assumption that clusters are standard candles), as well as on the
underlying cosmological parameters.  In particular, we find that, if
clusters are virialized objects, the cluster distribution in the
$\Delta S_{\nu}-T$ plane should be narrow. The predicted tightness of
the $\Delta S_{\nu}-T$ relation makes it especially useful for
quantifying clusters physical properties, possible deviations from
virialization, and to detect the presence of stochastic scatter in
($T,\Delta S_{\nu},\theta$), i.e. it is a useful tool to test whether
clusters can be used as ``standard candles''.
 
On the other hand, under the assumption that clusters can be used as
``standard candles'', we show that deformations of the fundamental
plane and of clusters scaling relations that involve $\Delta S_{\nu}$,
can be used to determine cosmological parameters. The constraints on
cosmological parameters so obtained are complementary to those
obtained e.g., from CMB primary anisotropies, cluster abundances, or
clusters central SZ decrements.

Our results show that the choice of statistic can have
a significant impact on the derived constraints, at the
level of a factor of several on the constrained parameters.

\section*{Acknowledgments}
We thank Joe Mohr and G. Holder for many stimulating discussions, and G. Bryan,
U. Seljak \& J. P. Hughes for useful comments. LV acknowledges NASA grant
NAG5-7154 and thanks Caltech, where part of this research was completed, for
hospitality. ZH is supported by NASA through the Hubble Fellowship grant
HF-01119.01-99A, awarded by the Space Telescope Science Institute, which is
operated by the Association of Universities for Research in Astronomy, Inc.,
for NASA under contract NAS 5-26555.

\end{document}